\documentclass[11pt]{article}%
\usepackage{amsfonts}
\usepackage{mathrsfs}
\usepackage{amsmath}
\usepackage{amssymb}
\usepackage{graphicx}
\usepackage{latexsym}
\usepackage{indentfirst}%
\hyphenpenalty=1500  \numberwithin{equation}{section}
\newtheorem{thm}{Theorem}[section]
\newtheorem{lema}[thm]{Lemma}
\newtheorem{prop}[thm]{Proposition}
\newtheorem{rmk}[thm]{Remark}
\newtheorem{coro}[thm]{Corollary}

\numberwithin{equation}{section}

\renewcommand{\a}{\alpha}

\def\a{\alpha}
\def\b{\beta}
\def\r{\gamma}

\def\P{\partial}
\def\p{\rho}
\def\R{{\bf R}}

\def\l{\lambda}

\def\pp{\tilde{\rho}}

\def\u{\tilde{u}}

\def\dis{\displaystyle}

\title{Global existence and $L^{p}$ convergence rates of planar waves for three-dimensional bipolar Euler-Poisson systems}
\author{Jie Liao$^{1}$, Yeping Li$^{2}$\thanks{Corresponding author. {\it E-mail address:
ypleemei@gmail.com}}, \\
$^1${\footnotesize \it Department of Mathematics, East China
University
of Science and Technology, Shanghai 200237, P. R. China.}\\
$^2${\footnotesize \it Department of Mathematics, East China
University of Science and Technology, Shanghai 200237, P. R.
China.}}

\date{}

\begin{document}

\maketitle

\noindent{\bf Abstract:} In the paper, we consider a
multi-dimensional bipolar hydrodynamic model from semiconductor
devices and plasmas. This system takes the form of Euler-Poisson
with electric field and frictional damping added to the momentum
equations. We show the global existence and $L^{p}$ convergence
rates of planar diffusion waves for multi-dimensional bipolar
Euler-Poisson systems when the initial data are near the planar
diffusive waves. A frequency decomposition and approximate Green
function based on delicate energy method are used to get the optimal
decay rates of the planar diffusion waves. To our knowledge, the
$L^p(p\in[2,+\infty])$-convergence rate of planar waves improves the
previous results about the $L^2$-convergence rates.\\

\noindent{\bf Key words:} Bipolar Euler-Poisson system, planar wave,
approximate Green function, smooth solution, energy estimates.\\

\noindent{\bf AMS subject classifications:} 35M20, 35Q35, 76W05.\\

\section{Introduction.}
In this paper, we consider the following bipolar Euler-poisson
system (hydrodynamic model) in three space dimension:
\begin{equation}\label{l1}
 \left\{
\begin{array}{lcr}
\partial_{t} \rho^{+}+ \mathrm{div}(\rho^{+} u^{+})=0,\vspace{2.5mm}\\
\partial_{t} (\rho^{+} u^{+}_{i})+\mathrm{div}(\rho^{+} u^{+}_{i}u^{+})+\partial_{x_{i}} P(\rho^{+})=-\rho^{+} u^{+}_{i} +\rho^{+}\partial_{x_{i}}\phi ,~~1\leq i\leq 3, \vspace{2.5mm}\\
\partial_{t} \rho^{-}+ \mathrm{div}(\rho^{-} u^{-})=0,\vspace{2.5mm}\\
\partial_{t} (\rho^{-} u^{-}_{i})+\mathrm{div}(\rho^{-} u^{-}_{i}u^{-})+\partial_{x_{i}} P(\rho^{-})=-\rho^{-} u^{-}_{i} - \rho^{-}\partial_{x_{i}}\phi ,~~1\leq i\leq 3, \vspace{2.5mm}\\
\Delta \phi = \rho^{+} -
\rho^{-},\lim_{|x|\rightarrow\infty}|\nabla\phi|=0,
\end{array}\right.
\end{equation}
with initial data
\begin{equation}\label{l2}
 (\rho^{\pm}, u^{\pm})(x,0)=(\rho_0^{\pm}(x), u_0^{\pm}(x)),
\end{equation}
where $\rho^{\pm}$ are the two particles's densities,
$\rho^{\pm}u^{\pm}= (\rho^{\pm}u^{\pm}_{1} , \rho^{\pm}u^{\pm}_{2},
\rho^{\pm}u^{\pm}_{3})$ are current densities,  $\phi$ is the
electrostatic potential, and $P(\rho^{\pm})$ are pressures. As
usual, we assume the pressure $P(\rho)$ be  smooth function in a
neighborhood of a constant state $\rho^*$ with $P^{\prime}(\rho)>0$.
The bipolar Euler-Poisson equations are generally used in the
description of charged particle fluids, for example, electrons and
holes in semiconductor devices, positively and negatively charged
ions in a plasma. This model takes an important role in the fields
of applied and computational mathematics, and we can see more
details in \cite{J,MRS,SM} etc..

Due to their physical importance, mathematical complexity and wide
range of applications, many efforts were made for the
multi-dimensional bipolar hydrodynamic equations from semiconductors
or plasmas. Li \cite{L2} showed existence and some limit analysis of
stationary solutions for the multi-dimensional bipolar Euler-Poisson
system. Ali and J\"{u}ngel \cite{AJ}, Li and Zhang \cite{LZ} and
Peng and Xu \cite{PX} studied the global smooth solutions of the
Cauchy problem for multidimensional bipolar hydrodynamic models in
the Sobolev space $H^l(\mathbb{R}^d)(l>1+\frac d2)$ and in the Besov
space, respectively. Ju \cite{J1} discussed the global existence of
smooth solutions to the initial boundary value problem for the
three-dimensional bipolar Euler-Poisson system. Li and Yang
\cite{LY} and Wu and Wang \cite{WW} showed global existence and
$L^2$ decay rate of the smooth solutions to the three dimensional
bipolar Euler-Poisson systems when the initial data are small
perturbation of the constant stationary solution. Huang, Mei and
Wang \cite{HMW} showed large time behavior of solution to
$n$-dimensional bipolar hydrodynamic model for semiconductors when
the initial data are near to the planar diffusion waves. Ali and
Chen \cite{AC} studied the zero-electron-mass limit in the
Euler-Poisson system for both well- and ill-prepared initial data.
Lattanzio \cite{L} and Li \cite{L1} investigated the relaxation
limit of the multi-dimensional bipolar isentropic Euler-Poisson
model for semiconductors, respectively. Ju, etc. \cite{JLLJ}
discussed the quasi-neutral limit of the two-fluid multi-dimensional
Euler-Poisson system. Moreover, it is worth to mentioning that there
are a lot of reference about the one-dimensional bipolar
Euler-Poisson equation, and the interesting reader can refer to
\cite{DMRS,GM,GHL,HL,HMWY,HZ1,HZ2,N1,T,ZH,ZL} and the reference
therein. In particular, motivation by \cite{GM,[H-L]}, Gasser, Hsiao
and Li \cite{GHL} found that the frictional damping is the key to
the nonlinear diffusive phenomena of hyperbolic waves, and
investigated the diffusion wave phenomena of smooth ``small"
solutions for the one-dimensional bipolar hydrodynamic model. Huang
and Li \cite{HL} also studied the large-time behavior and
quasi-neutral limit of $L^\infty$ solution of the one-dimensional
Euler-Poisson equations for large initial data with vacuum. That is,
they showed that the weak entropy solution of the one-dimensional
bipolar Euler-Poisson system converges to the nonlinear diffusion
waves. Then Huang, Mei and Wang \cite{HMW} showed the planar
diffusive wave stability to $n(n\geq2)$-dimensional bipolar
hydrodynamic model for semiconductors, and obtained the optimal
$L^2$ and $L^\infty$ decay rates. In this paper, we are going to
reconsider global existence of the smooth solution for the
multi-dimensional bipolar Euler-Poisson systems, in particular, we
try to establish the $L^{p}(p\in[2,+\infty])$ convergence rates of
planar waves.

In the following discussion, we assume that the initial data are a
small perturbation of the diffusion profile constructed later with
small wave strength. Let the initial data $\rho_0^{\pm}(x)$ be
strictly positive and satisfy
$$
\lim_{x_1\rightarrow \pm\infty}\rho_0^{\pm}(x)=\rho_\pm, %\eqno{(1.3)}
$$
where $ \rho_\pm>0$ are two far field constants with $\rho_-\ne
\rho_+$. Similar as the consideration of planar diffusion waves of
damped Euler equations in  \cite{[W-Y3],lwy09}, to define the
multi-dimensional planar diffusion wave, we first consider the one
dimensional diffusion equation
\begin{equation}\label{p5}
 \partial_t w= P(w)_{x_1x_1},
\end{equation}
which can be derived from the bipolar Euler-Poisson equations with
the relaxation terms in one dimensional case by imposing the Darcy's
law, cf. \cite{GHL,GM}. Then a multi-dimensional diffusion wave
$w(x, t)$ is a one dimensional profile in multi-dimensional space.
That is, $w(x,t)=W(x_1/\sqrt{1+t})$ is a self-similar solution of
the equation (\ref{p5}) connecting two end states $\rho_\pm$  at
$x_1=\pm\infty$. Denote $\zeta=\frac{x_1}{\sqrt{t+1}}$, then
$W(\zeta)$ satisfies
$$
-\frac 12 \zeta\partial_\zeta
W=\partial_\zeta(P'(W(\zeta))\partial_\zeta W).
$$

For simplicity, let the initial velocity $u_0^{\pm}(x)$ vanish as
$x_1\rightarrow \pm\infty$, that is,
$$
\lim_{x_1\rightarrow \pm\infty}u^{\pm}_0(x)=0,
$$
which  implies that there is no mass flux coming  in from
$x_1=\pm\infty$. This assumption could be removed in a technical way
similar to the argument for one dimensional problem because  the
momentum at $x_1=\pm\infty$ decays exponentially induced by the
linear relaxation terms.

We now recall the bipolar Euler-Poisson systems (\ref{l1}) in one space dimension:
\begin{equation}\label{l3}
 \left\{
\begin{array}{lcr}
\partial_{t} \rho^{+}+ \partial_{x_{1}}(\rho^{+} u_{1}^{+})=0,\vspace{2.5mm}\\
\partial_{t} (\rho^{+} u^{+}_{1})+ \partial_{x_{1}}(\rho^{+} u^{+}_{1}u^{+}_{1})+\partial_{x_{1}} P(\rho^{+})=-\rho^{+} u^{+}_{1} +\rho^{+}E , \vspace{2.5mm}\\
\partial_{t} \rho^{-}+ \partial_{x_{1}}(\p^{-} u^{-}_{1})=0,\vspace{2.5mm}\\
\partial_{t} (\p^{-} u^{-}_{1})+\partial_{x_{1}}(\p^{-} u^{-}_{1}u^{-}_{1})+\partial_{x_{1}} P(\p^{-})=-\p^{-} u^{-}_{1} - \p^{-}E , \vspace{2.5mm}\\
\partial_{x_{1}}E = \p^{+} - \p^{-},\lim_{x_1\rightarrow-\infty}E(x_1,t)=0.
\end{array}\right.
\end{equation}
Denote the solution of (\ref{l3}) by $ (\pp^{\pm}, \u_1^{\pm},\tilde
{E})(x_1,t)$. When
$$
\lim\limits_{x_1\rightarrow\pm\infty} \pp^{\pm}(x_1, 0)=\p_{\pm},\ \
\lim\limits_{x_1\rightarrow\pm\infty} \tilde{u}^{\pm}_1(x_1, 0)=0,
$$
the time-asymptotic behavior of $(\pp^+,\u_1^+,\pp^-,
\u_1^-)(x_1,t)$ has been studied in \cite{GHL}, which is shown to be
a nonlinear diffusion profile governed by Darcy's law. Roughly
speaking, the solution $\pp^{\pm}(x_1, t)$ converge to a same
diffusion wave $W(x_1/\sqrt{1+t})$ up to a constant shift in $x_1$.
Note that more detailed assumptions on the initial data of the
one-dimensional problem \eqref{l3} will be specified in Theorem 2.1.

In this paper, we will generalize this time asymptotic behavior
towards a planar diffusion wave to three-dimensional case and
establish the related $L^p~(2\leq p\leq\infty)$ convergence rates.

As in the consideration of planar diffusion waves of damped Euler
equation in \cite{[W-Y3],lwy09} and of the bipolar Euler-Poisson
system in  \cite{HMW}, we do not directly compare the solution of
the problem (\ref{l1}) with the diffusion wave $W(x_1/\sqrt{1+t})$,
instead, we will compare it with the solution of one dimensional
problem (\ref{l3}). For this, without loss of generality,  let us
first assume the initial density $\pp^{\pm}(x_1, 0)$ in (\ref{l3})
satisfy
\begin{equation}\label{a1}
\int^{+\infty}_{-\infty}(\pp^{\pm}(x_1 ,0)-W(x_1))dx_1=0.
\end{equation}
For the multi-dimensional problem, the shift function $\delta_0(x')$
, where we used the notation $x'=(x_{2},x_{3})$, can be chosen as in
\cite{[W-Y3],lwy09} such that the initial density function satisfies
$$
\int^{+\infty}_{-\infty}(\p^{\pm}(x,
0)-W(x_1+\delta_0^{\pm}(x')))dx_1=0.%\eqno{(1.7)}
$$
Note that $\delta_0^{\pm}(x')$ is then uniquely determined by
$$
\delta_0^{\pm}(x') = \frac{1}{\rho_+ - \rho_-}\int_{-\infty}^{\infty}
(\rho^{\pm}(x,0)-W(x_1))dx_1,
$$
for $\rho_-\ne \rho_+$. Moreover, we  assume that
basically the shift is uniform in directions other than $x_1$
at infinity, that is,
$$
\lim_{|x'|\rightarrow+\infty}\frac{1}{\rho_+ -
\rho_-}\int^{+\infty}_{-\infty}(\p^{\pm}(x,
0)-W(x_1))dx_1=\delta_*^{\pm},%\eqno{(1.8)}
$$
Note that this assumption simplifies the problem and it remains
unsolved for general perturbation when this assumption fails. An
immediate consequence of this assumption is that
$$
\lim_{|x'|\rightarrow+\infty}\delta_0^{\pm}(x')=\delta_*^{\pm}.
$$
And for simplicity, we assume $\delta_*^{\pm}=\delta_*$ be same constants.
With these notations, the main purpose here is to show that the
solutions $(\p^{\pm}, u^{\pm})$ of (\ref{l1}) converge to $(\bar{\p}^{\pm},
\bar{u}^{\pm})$ with certain time decay rates, where
\begin{equation}\label{l4}
\left\{
\begin{array}{lcr}
\bar{\p}^{\pm}(x,t)=\pp^{\pm} (x_1+\delta(x', t), t),\vspace{2.5mm}\\
\bar{u}^{\pm}(x,t)=(\tilde{u}^{\pm}_1(x_1+\delta(x', t), t), 0 ,0),\vspace{2.5mm}\\
%\bar{u}^{\pm}_i(x,t)=&0,~\hspace{5mm}~i=2,3,\vspace{2.5mm}\\
\bar E  (x,t)=(\tilde E(x_1+\delta(x', t), t), 0 ,0)   ,\vspace{2.5mm}\\
\delta(x', t)=\delta_*+e^{- t}(\delta_0(x')-\delta_*),
\end{array}
\right.
\end{equation}
in which $\pp^{\pm},\tilde{u}^{\pm}_1$ and  $\tilde E$ are solution of
(\ref{l3}). In the following discussion, we will also assume that
the shift generated by the initial data satisfies
\begin{equation}\label{l11}
  |\partial_{x'}^\beta(\delta_0^{\pm}(x')-\delta_*) |\le
C(1+|x'|^2)^{-N},
\end{equation}
for any multi-index $\beta$ and any positive integers  $N$. Here,
$C$ is a constant depending only  on $\beta$. This assumption
implies that the shift $\delta_0(x')$ decays to $\delta_*$ almost
exponentially. Again, this assumption can be reduced to the
constraint on the initial perturbation. More precise construction of
the background planar diffusion wave $(\bar{\rho}^{\pm},
\bar{u}^{\pm})(x,t)$ and its properties will be given in Lemma 2.3
below.

%Before stating the main result, we need a few more notations.
Throughout this paper, we denote any generic constant by $C$. The
usual Sobolev space is denoted by $W^{s,p}({\R}^n)$, $s\in {\bf
Z_+}$, $p\in
 [1,\infty]$
 with the norm
$$
\|f\|_{W^{s,p}} :=\sum^s_{|\a|=0}\|\partial^\a f\|_{L^p},
$$
where $\P^\a$ used for $\P^\a_x$ without confusion. In particular,
$W^{s,2}({\R}^n)=H^s({\R}^n)$. Set
\begin{equation}\label{l5}
 \begin{array}{lcr}
 V^{\pm}(x,t)=\p^{\pm}(x,t)-\bar{\p}^{\pm}(x,t), \vspace{2.5mm}\\
 U^{\pm}(x,t)=(u_1^{\pm}(x,t)-\bar{u}^{\pm}_1(x,t), u_2^{\pm}(x,t), u_3^{\pm}(x,t))
   , \vspace{2.5mm}\\
 \nabla \varphi =  \nabla \phi -\bar E ~(~\rm{Note}~\mathrm{div}(\nabla \phi -\bar E )=0), \vspace{2.5mm}\\
 K(x,t) = V^{+}(x,t) - V^{-}(x,t),
%  \vspace{2.5mm}\\
% X(x,t) = V^{+}(x,t) + V^{-}(x,t),
 \end{array}
\end{equation}
and also denote
\begin{equation}\label{l6}
 \nu^{\pm} (x,0)=\int^{x_1}_{-\infty}V^{\pm}(x_1,x', 0)dx_1, ~~~
\nu^{\pm}_t(x,0)=\int^{x_1}_{-\infty}V^{\pm}_t(x_1, x', 0)dx_1.
\end{equation}
Note that the time derivative on the initial data  can be
 defined by the compatibility of the
initial data through the equation (\ref{l7}).

Now we state the main results in this paper. Note that we consider
only the spatial dimension $n=3$ in this paper. However, we will
still use notation $n$ in the below theorem for convenience to
extend our result to other high dimensional  cases, since other
higher dimensional cases can be considered similarly.

\begin{thm}
Let $(\bar{\rho}^{\pm}, \bar{u}^{\pm})(x,t)$  in (\ref{l4}) be
planar diffusion waves with a shift $\delta(x', t)$ constructed
above. (See more precisely its properties in Lemma 2.3.)
 For $k\geq 4$, assume that the initial data $(\rho^{\pm}, u^{\pm})(x,0)$
satisfy the smallness assumption
%\begin{equation}\label{l8}
 $$
 \begin{array}{rl}
 |\p_+-\p_- |+ \|(\nu^{\pm}, \nu^{\pm}_t)(\cdot, 0) \|_{L^2\cap
L^1}+\|(\p^{\pm}-\bar{\p}^{\pm})(\cdot,0) \|_{L^1} +
\|(\p^{\pm}-\bar{\p}^{\pm}, u^{\pm}-\bar{u}^{\pm})(\cdot, 0)
\|_{H^k}   \leq \epsilon_0,
\end{array}
$$%\end{equation}
where $\epsilon_0>0$ is a sufficiently small constant. Then  \\
(i) (Global existence) There exist unique global classical solution $(\rho^{\pm},  u^{\pm}, \nabla\phi)$  to the system (\ref{l1})-(\ref{l2})  that
$$V^{\pm}(t,x),  U^{\pm}(t,x)\in C([0,\infty),H^{k}({\R}^n))\cap C^1((0,\infty), H^{k-1}({\R}^n)), \hspace{2mm}
\nabla\phi \in W^{k,6}({\R}^n).$$
(ii) ($L^{p}$ convergence) Moreover, for $|\r|\leq k-2$, $p\in[2, \infty]$, we have
$$
 \begin{array}{rl}
 \|\P^\r_xV^{\pm} \|_{L^p} \leq &C \epsilon_0 (1+t)^{-\frac{n}{2}(1-\frac{1}{p})-\frac{|\r|+1}{2}}, \vspace{2.5mm}\\
 \|\P^\r_xU^{\pm} \|_{L^p} \leq
&C \epsilon_0 (1+t)^{-\frac{n}{2}(1-\frac{1}{p})-\frac{|\r|+2}{2}}.
\end{array}
$$
(iii) (Estimates on $\varphi$ and $K=V^{+}-V^{-}$)  For $|\gamma|\leq k-2$,
$$
 \|\P^\gamma_x K \|_{L^2} \leq C \epsilon_0 (1+t)^{-{5\over 4}n-2- {|\gamma|\over 2}}
 ,  \hspace{3mm}
 \|\P^\gamma_x \nabla\varphi \|_{L^6} \leq   C \epsilon_0 (1+t)^{-{5\over 4}n-2- {|\gamma|\over 2}}   ,
$$
and for $|\gamma|= k-1$,
$$
 \|\P^\gamma_x K \|_{L^2} \leq C \epsilon_0 (1+t)^{-{5\over 4}n-1- {k\over 2}}
 ,  \hspace{3mm}
 \|\P^\gamma_x \nabla\varphi \|_{L^6} \leq   C \epsilon_0 (1+t)^{-{5\over 4}n-1- {k\over 2}}  .
$$
\end{thm}

\begin{rmk}
As noted in \cite{lwy09}, in general, if the shift of the profile is
not exactly captured, the decay rates for $V^{\pm}$ and $U^{\pm}$
should be $\frac 12$ lower than the one given in the above theorem
even in one space dimensional case. Here, the reason that the above
decay estimate holds is that the shift due to the initial
perturbation introduced above so that when we apply the Green
function, the term corresponding to the initial data yields an extra
$(1+t)^{-\frac 12}$ decay after taking the anti-derivative of the
initial perturbation. Moreover, under the condition (\ref{l11}) on
the initial shift, even though the anti-derivative of the
perturbation can not be defined for all time as the shift function
is not precisely defined, we know that $\delta_*$ is exactly the
final shift  when $t$ tends to infinity of the profile because the
initial perturbation will spread out eventually.
\end{rmk}

\begin{rmk}
We note here that the $L^{2}$ decay rates of $K$ is higher than
that of $V^{\pm}$. However, we can only get the decay estimates on
derivatives of $K$ up to  $(k-1)$-th order.
\end{rmk}

\begin{rmk}
Compared with \cite{LY, WW}, our initial data are the small
perturbation of the planar waves, instead of the constant states. In
the meanwhile, here we can show the $L^p(p\in[2,+\infty])$
convergence rates of the planar waves of the three-dimensional
bipolar Euler-Poisson equations. This  improved the results in
\cite{HMW}. Moreover, here we only consider the case that the far
fields of two particles' velocity in the $x_1$-direction are same,
see (\ref{a1}), namely, the switch-off case. However, we believe
that the same results also hold for the switch-on case. Indeed,
using the gap function with exponential decay in \cite{HMW}, we can
show the similar results for the switch-on case.
\end{rmk}

The outline of the proof of the main theorems is as follows. First,
we notice that the equations for $V^{\pm}$ are coupled by
$\nabla\varphi$, which is expressed by nonlocal  Riesz potential
$\nabla \varphi = \nabla \Delta^{-1} K$, with $K=V^+-V^-$. So we
need to have some good estimates on $K$ before the estimate of
$V^{+}$ and $V^{-}$. Luckily we note that $K$ satisfies the damping
``Klein-Gordon" type equations with an addition good term to perform
the energy estimate. The estimates of $K$ and $\nabla\varphi$ are
given in Section 2, where the algebraic decay rates of $K$ in the
$L^2$-norm are derived by some delicate energy methods, which will
be used to obtain the $L^p(p\in[2,+\infty])$-convergence rates of
the solutions in the subsequent. Next, we use the frequency
decomposition based energy method introduced in \cite{lwy09}, which
combines the approximation Green function and energy method, to
prove global existence and $L^{p}$ convergence results, see (i) and
(ii) in Theorem 1.1. This method captures the low frequency
component in the approximate Green function and avoids the
singularity in the high frequency component. That is, we firstly
show the precise algebraic decay estimate of $V^{\pm}$ in the low
frequency component, which dominant the decay of the perturbation,
and then obtain the better decay rates of the high frequency
component of $V^{\pm}$ in the $L^2$-norm by energy methods. For the
high frequency component, one has an additional  $\rm
Poincar\acute{e}$-type inequality to close the energy estimate. Note
that the lack of $\rm Poincar\acute{e}$ inequality in the whole
space is usually the essential difficulty in the energy estimate
which is in contrast to the problem in a torus. This in some sense
illustrates the essence of the Green function on the decay rate
related to the frequency.

The rest of the paper is arranged as follows. In Section 2, we will
reformulate the system around  a planar diffusion wave defined in
(\ref{p5}) and then state some known properties of this background
diffusion wave. In Section 3, we will study the energy estimate of
$K=V^{+}-V^{-}$ and prove part (iii) in Theorem 1.1. The frequency
decomposition based energy method will be carried out in Section 4
and 5, where in Section 4, we will study the approximate Green
function and then the main $L^p$ estimates on the low frequency
component of $V^{\pm}$,  and in Section 5, we will study the $L^2$
energy estimates on  the high frequency component of $V^{\pm}$.
Finally, we will complete the proof to part (i) and (ii) of Theorem
1.1 in Section 6.

\section{Preliminaries.}
In this section, we will first derive the equations
for the perturbation functions $V^{\pm}$ and $U^{\pm}$ defined in
(\ref{l5}). Then we will recall some results on the background
diffusion waves.

\subsection{Reduced system.}
We first derive the system for the perturbation of the nonlinear
planar diffusion wave. Then, from (\ref{l1})  and (\ref{l3}), we
have the equations for $V^{\pm}$ that
\begin{equation}\label{l7}
V^{\pm}_{t}+(\bar{\p}^{\pm}+V^{\pm}){\rm div}U ^{\pm}
= R_{\rho^{\pm}}-(U^{\pm}\cdot\nabla)(\bar{\rho}^{\pm}+V^{\pm})
-V^{\pm}(\bar{u}_1^{\pm})_{x_1}-\bar{u}_1^{\pm}V^{\pm}_{x_1},
\end{equation}
where
$$
R_\rho^{\pm}=  [-\bar{\p^{\pm}}(x, t)\delta^{\pm}_t(x', t)]_{x_1}.
$$
Similarly, the equations for $U_1^{\pm}$ are
$$
(U_1^{\pm})_t+(\bar{\p}^{\pm}+V^{\pm})^{-1}[P(\bar{\p}^{\pm}+V^{\pm})-P(\bar{\p}^{\pm})]_{x_1}+
U_1^{\pm} =
\frac{P(\bar{\p}^{\pm})_{x_1}V^{\pm}}{\bar{\p}^{\pm}(\bar{\p}^{\pm}+V^{\pm})}
+R_{u_1^{\pm}}-R_1^{\pm} \pm \partial_{x_{1}} \varphi,
$$
and for $i= 2, 3$,
$$
(U_i^{\pm})_t+(\bar{\p}^{\pm}+V^{\pm})^{-1}[P(\bar{\p}^{\pm}+V^{\pm})-P(\bar{\p}^{\pm})]_{x_i}+
U_i^{\pm}
=\frac{P(\bar{\p}^{\pm})_{x_i}V^{\pm}}{\bar{\p}^{\pm}(\bar{\p}^{\pm}+V^{\pm})}-
(\bar{\p}^{\pm})^{-1}P(\bar{\p}^{\pm})_{x_i}-R_i^{\pm} \pm
\partial_{x_{i}} \varphi,
$$
where
$$
\begin{array}{rl}
R_{u_1^{\pm}}=&-[\tilde{u}_1^{\pm}(x_1+\delta(x', t), t)\delta_t(x', t)]_{x_1},\vspace{2.5mm}\\
R_1^{\pm}=&U^{\pm}\cdot\nabla(\bar{u}_1^{\pm}+U_1^{\pm})+\bar{u}_1^{\pm}(U_1^{\pm})_{x_1},\vspace{2.5mm}\\
R_i^{\pm}=&U^{\pm}\cdot\nabla U_i^{\pm}+\bar{u}_1^{\pm}(U_i^{\pm})_{x_1},~ 2\leq i\leq n.
\end{array}
$$
The equation for $\varphi$ is simply
$$
\Delta \varphi = V^{+} - V^{-} =K,
$$
it is directly that the perturbed electric field $\nabla \varphi$ can be expressed by the Riesz potential as a nonlocal term
\begin{equation}\label{l12}
\nabla \varphi = \nabla \Delta^{-1} K.
\end{equation}
Then the system  for the perturbation  $(V^{\pm}, U^{\pm}, \varphi)$ can be summarized as
\begin{equation}\label{l13}
 \left\{
\begin{array}{lcr}
V^{\pm}_{t}+(\bar{\p}^{\pm}+V^{\pm}){\rm div}U^{\pm}=Q^{\pm},\vspace{2.5mm}\\
(U_i^{\pm})_t+(\bar{\p}^{\pm}+V^{\pm})^{-1}({\mathcal P}(V^{\pm},
\bar{\p}^{\pm})V^{\pm})_{x_i}+ U_i^{\pm}=H^{\pm}_i \pm
\partial_{x_{i}}\varphi, \quad 1\le i\le 3,
\end{array}
\right.
\end{equation}
where ${\mathcal P}(V^{\pm}, \bar{\p}^{\pm})=\dis{\int^1_0 }P^\prime(\bar{\p}^{\pm}+\theta V^{\pm})d\theta$, and
$$
\begin{array}{rl}
Q^{\pm}=&R_{\rho^{\pm}}-(U^{\pm}\cdot\nabla)(\bar{\rho}^{\pm}+V^{\pm})-V^{\pm}(\bar{u}^{\pm}_1)_{x_1}
-(\bar{u}^{\pm}_1)V^{\pm}_{x_1},\vspace{2.5mm}\\
H_1^{\pm}=&R_{u^{\pm}_1}+\frac{P(\bar{\p}^{\pm})_{x_1}V^{\pm}}{\bar{\p}^{\pm}(\bar{\p}^{\pm}+V^{\pm})}-R_1^{\pm},\vspace{2.5mm}\\
H_i^{\pm}=&-\frac{P(\bar{\p}^{\pm})_{x_i}}{\bar{\p}^{\pm}} +\frac{P(\bar{\p}^{\pm})_{x_i}
V^{\pm}}{\bar{\p}^{\pm}(\bar{\p}^{\pm}+V^{\pm})}-R_i^{\pm},  ~~2\le i\le 3.
\end{array}
$$
Moreover, we can deduce the equation for  $V^{\pm}(x,t)$  from (\ref{l13}) as
\begin{equation}\label{l14}
 V^{\pm}_{tt}-\triangle[{\mathcal P}(V^{\pm}, \bar{\p}^{\pm})V^{\pm}]+ V^{\pm}_{t}=\tilde{Q}(V^{\pm}, U^{\pm},\bar{\p}^{\pm}, \bar{u}^{\pm}_1)
\mp \mathrm{div}[( \bar{\p}^{\pm}+V^{\pm} )\nabla\varphi],
\end{equation}
where
$$
 \begin{array}{rl}
 &\tilde{Q}(V^{\pm},U^{\pm}, \bar{\p}^{\pm}, \bar{u}^{\pm}) \vspace{2.5mm}\\
 = &[(R_{\rho^{\pm}})_t+R_{\rho^{\pm}} ]-(1
+\partial_t)(V^{\pm}\bar{u}^{\pm}_1)_{x_1}
-\mathrm{div}[(\bar{\p}^{\pm}+V^{\pm})_tU^{\pm}]-\mathrm{div}[(\bar{\p}^{\pm}+V^{\pm})H^{\pm}],
\end{array}
$$
with $H^{\pm}=(H_1^{\pm}, \cdots, H_3^{\pm})$. By linearizing (\ref{l14}) around $\bar{\p}$,
we have
\begin{equation}\label{l15}
 \begin{array}{rl}
&V^{\pm}_{tt}-\triangle(a^{\pm}(x,t)V^{\pm})+  V^{\pm}_{t}\vspace{2.5mm}\\
=&\tilde{Q}(V^{\pm}, U^{\pm},\bar{\p}^{\pm}, \bar{u}^{\pm}_1)
+\triangle  ({\mathcal P}_1(\bar{\p}, V)V^2) \mp \mathrm{div}[(
\bar{\p}^{\pm}+V^{\pm} )\nabla\varphi]
\vspace{2.5mm}\\
=: &F^{\pm} \mp \mathrm{div}[( \bar{\p}^{\pm}+V^{\pm}
)\nabla\varphi],
\end{array}
\end{equation}
where $a^{\pm}(x,t)=P^\prime(\bar{\p}^{\pm})$ and
$$
{\mathcal P}_1(\bar{\p}^{\pm}, V^{\pm})=\int^1_0 (\int^{\theta_1}_0
P^{\prime\prime}(\bar{\p}^{\pm}+\theta_2V^{\pm})d\theta_2 ) d\theta_1.
$$
Since
$$
(R_{\rho^{\pm}})_t+R_{\rho^{\pm}}= (-\bar{\p}^{\pm}_t(x,
t)\delta_t(x^\prime, t) )_{x_1},
$$
direct calculation shows that $F^{\pm}=F(V^{\pm}, U^{\pm}, \bar{\p}^{\pm},\bar{u}^{\pm})$ is in
divergence form, that is,
\begin{equation}\label{l18}
F^{\pm}=\sum (F^{\pm,i})_{x_i}+\sum (F^{\pm,ij})_{x_ix_j},
\end{equation}
where, without confusion, we omit the $\pm$ sign,
$$
\begin{array}{rl}
F^1=&-\bar{\p}_t\delta_t-(\bar{\rho}\bar{u}_1\delta_t)_{x_1}, \ \ \ F^i=-P(\bar{\p})_{x_i}, \,\, 2\le i\le n, \vspace{2.5mm}\\
F^{11}=& \bar{\p}(2\bar{u}_1U_1+U^2_1)+V(\bar{u}_1+
U_1)^2+{\mathcal P}_1(\bar{\p}, V)V^2,\vspace{2.5mm}\\
F^{1i}=&F^{i1}=2 [(\bar{\p}+V)(\bar{u}_1+U_1)U_i ],\,\, 2\le i\le n, \vspace{2.5mm}\\
F^{ij}=&(\bar{\p}+V)U_iU_j+\delta_{ij}{\mathcal P}_1(\bar{\p},
V)V^2,\,\, 2\le \ i,j\le n.
\end{array}
$$
Here $\delta_{ij}$ is the Kronecker symbol. On the other hand, by
linearizing (\ref{l13})$_2$ around $\bar{u}^{\pm}$, we have
\begin{equation}\label{l16}
 U^{\pm}_t + (\bar{\p}^{\pm})^{-1}\nabla(a^{\pm}(x,t) V^{\pm}) + U^{\pm}=\bar{H}^{\pm} \pm \nabla\varphi ,
\end{equation}
where, again without confusion, we omit  the $\pm$ sign,
$$
\begin{array}{rl}
\bar{H}_1=&  R_u-\bar{\p}^{-1}[{\mathcal P}_1(\bar{\p}, V)V^2]_x
-\frac{P(\bar{\p}+V)_xV}{\bar{\p}(\bar{\p}+V)}-R_1, \vspace{2.5mm}\\
\bar{H}_i=&-\frac{P(\bar{\p})_{x_i}}{\bar{\p}}-\frac{P(\bar{\p}+V)_{x_i}V}{\bar{\p}(\bar{\p}+V)}-\bar{\p}^{-1}[{\mathcal
P}_1(\bar{\p}, V)V^2]_{x_i}-R_i,~~ 2\le i\le n.
\end{array}
%\eqno{(2.13)}
$$

\subsection{Background profile.}
For later use, we include the following known estimates on the
background planar  wave, cf. \cite{[W-Y3]}. By the definition of
$W(x_1)$, we know for any integer $N$,
$$
\begin{array}{rl}
\displaystyle{\sup_{x_1>0}|W (x_1)-\p_+|}+&\displaystyle{\sup_{x_1<0}}|W (x_1)-\p_-|\leq
C|\p_+-\p_-|(1+x_1^2)^{-N},\vspace{2.5mm}\\
 |\P^h_{x_1}W (x_1)|\leq &C|\p_+-\p_-|(1+x_1^2)^{-N},\ \ (h>0).
\end{array}
$$
Recall that we have assumed  in (\ref{l11}) for any multi-index $\beta$,
$$
|\P^{\beta}_{x'}(\delta_0(x')-\delta_*)|\leq C(1+|x'|^2)^{-N}.
$$
First, let us recall the results about the one-dimensional bipolar
Euler-Poisson system (\ref{l3}).
\begin{thm} (see \cite{GHL})
Let
$(\tilde{\p}^\pm, \tilde{u}_1^\pm)(x_1, 0)$ be the initial data of
one-dimensional bipolar Euler-Poisson system (\ref{l3}) and fix an
integer $m\geq 2$. If there exists a small positive constant $E_\p$
such that the initial data $(\tilde{\p}^\pm, \tilde{u}_1^\pm)(x_1,
0)$ satisfy
$$
\begin{array}{rl}
&\dis{ |\p_+-\p_- |+ \|\int^{x_1}_{-\infty}(\tilde{\p}^{\pm}(z, 0)-W  (z))dz \|_{L^2}}\vspace{2.5mm}\\
&+ \|\tilde{\p}^{\pm}(\cdot, 0)-W  (\cdot) \|_{H^{m+1}}
+ \|\tilde{u}^{\pm}_1(\cdot, 0)-\psi(\cdot, 0) \|_{H^{m+1}}\leq E_\p,
\end{array}
%\eqno{(2.14)}
$$
and
$$\int^{x_1}_{-\infty}(\tilde{\p}^{\pm}(z,
0)-W(z))dz,\tilde{u}^\pm+P(W)_{x_1}\in L^1(\R),
$$
 then  (\ref{l3}) has global classical solution $(\tilde{\rho}^\pm,\tilde{u}^\pm, E)$ with
\begin{eqnarray*}
 &&\|\partial^k(\tilde{\rho}^\pm-W)(x_1,t) \|_{L^p(\R^1_{x_1})} \leq
CE_\p(1+t)^{-\frac12(1-\frac1p)-\frac{k+1}{2}},\vspace{2.5mm}\\
 &&\|\partial^k(\tilde{u}^\pm+P(W_{x_1}))(x_1,t) \|_{L^p(\R^1_{x_1})} \leq
CE_\p(1+t)^{-\frac12(1-\frac1p)-\frac{k+2}{2}}
\end{eqnarray*}
for any integer $k\leq m+1$ if $p=2$, and $k\leq m$ if
$p=\infty$. Moreover, there exists a positive constant $\beta$ such
that
$$
\|(\tilde{\rho}^{+} -
\tilde{\rho}^{-},\tilde{E})\|_{H^{m}(\R^1_{x_1})} \leq CE_\p
e^{-\beta t}.
$$
\end{thm}

\begin{rmk}
Note that $(\tilde{\p}^\pm, \tilde{u}_1^\pm)$ is an intermediate
state we constructed to approximate the one-dimensional diffusion
wave $W$. The assumptions on the initial data $(\tilde{\p}^\pm,
\tilde{u}_1^\pm)(x_1, 0)$, with $\tilde{\p}^\pm(x_1, 0)$ connecting
the two end states $\rho_\pm$, can be more regular than the
assumptions on the initial date of the original problem \eqref{l1}.
\end{rmk}

Next, from the definition of the planar diffusion waves
$(\bar\rho^{\pm}, \bar u^{\pm})$  in (\ref{l4}), we can readily have

\begin{lema} Under the assumptions in Theorem 2.1, the planar diffusion
waves $(\bar\rho^{\pm}, \bar u^{\pm})$ defined in (\ref{l4}) satisfy
$$
\begin{array}{rl}
\displaystyle{ \sup_{x^\prime} \|\P^\a(\bar{\p}^{\pm}_{x_1},
\bar{u}^{\pm}_1)(\cdot, x', t) \|_{L^2(\R^1_{x_1})}} \leq &
CE_\p(1+t)^{-{1+|\a|\over2}- {1\over4}},\vspace{2.5mm}\\
  \|\P^\a(\bar{\p}^{\pm}_{x_1},
\bar{u}^{\pm}_1)(\cdot, t) \|_{L^\infty(\R^n)} \leq &
CE_\p(1+t)^{-{1+|\a|\over2}}
\end{array}
%\eqno{(2.15)}
$$ for any multi-index $\a$ with $|\a|\leq m-1$, and
$$
\|\bar \p^{+} - \bar \p^{-}\|_{H^{m}(\R^1_{x_1})} \leq CE_\p
e^{-\beta t}.
$$
In addition, for $2\le i\le n$,
$$
\begin{array}{rl}
  \|\P^\a(\bar{\p}^{\pm}_{x_i},
(\bar{u}^{\pm}_1)_{x_i})(t) \|_{L^2(\rm \R^n)}\leq CE_\p e^{- t},\vspace{2.5mm}\\[2.5mm]
 \|\P^\a(\bar{\p}^{\pm}_{x_i},
(\bar{u}^{\pm}_1)_{x_i})(t) \|_{L^\infty(\rm \R^n)}\leq CE_\p e^{-
t}.
\end{array}
%\eqno{(2.16)}
$$
\end{lema}

Note here again that we can increase the regularity of the
assumptions on initial date of the one-dimensional problem to get
sufficient estimates on the planar diffusion wave.

\section{Estimates on $K=V^{+} - V^{-}$. }
In this section, we mainly give the estimate of $K=V^+-V^-$. Recall
the linearized equation (\ref{l15}) for $V^{\pm}$, we see that they
are coupled by $\nabla\varphi$, which is expressed by the Riesz
potential as in (\ref{l12}). i.e., $ \nabla \varphi = \nabla
\Delta^{-1} K$. So we need to have some good estimates on $K$, thus
$\nabla\varphi$, before the estimate of $V^{+}$ and $V^{-}$. To
begin with, we give a lemma on the relation of  $ \nabla \varphi $
and $K$.
\begin{lema}
If $K\in H^{l}(\R^n)$ for any integer $l>1$, then $\nabla \varphi
\in W^{l,6}(\R^n)$.
\end{lema}

\noindent{\bf Proof.} Note that $\nabla\varphi$  be expressed by the
Riesz potential
  $$  \nabla \varphi = \nabla \Delta^{-1} K = \mathcal R*K,$$
where $\hat {\mathcal R} = |2\pi \xi|^{-1}$ thus $\mathcal R= {1\over |x|^{n-1}}$. Here
$n=3$ is the space dimension. Then by Hardy-Littlewood-Sobolev
inequality \cite{ST} we have
\begin{equation}\label{l21}
  \|\nabla \varphi \|_{L^{6}} = \|  \mathcal R *K \|_{L^{6}} \leq C \| K\|_{L^{2}},
\end{equation}
and similarly, for any multi index $|\r|\leq k$,
\begin{equation}\label{l22}
  \|\P^{\r}\nabla \varphi \|_{L^{6}} = \|  \mathcal R * \P^{\r} K \|_{L^{6}} \leq C \| \P^{\r}K\|_{L^{2}},
\end{equation}
that is, $\nabla \varphi \in W^{l,6}(\R^n)$ if $K\in H^{l}(\R^n)$.

\begin{rmk}
By Sobolev injection, this lemma automatically indicates
$$\nabla \varphi \in L^{\infty}(\R^n).$$
\end{rmk}

Now we start to estimate $K$. The equation for $K$, from (\ref{l15}), is
$$
 K_{tt} - \Delta(a^{+}K) + K_{t} = F^{+}-F^{-}
 -\mathrm{div}[(\bar\p^{+} +\bar\p^{-} +V^{+} +V^{-} )\nabla\varphi]
 + \Delta[(a^{+}-a^{-})V^{-}] ,
$$
where $F^{\pm}$ on the right hand side are defined in (\ref{l18}). This equation can also be written as
\begin{equation}\label{l17}
 \begin{array}{rl}
&K_{tt} - \Delta(a^{+}K) + K_{t}  + (\bar\p^{+} +\bar\p^{-} +V^{+} +V^{-} )K
\vspace{2mm}\\
 = & F^{+}-F^{-}  - \nabla (\bar\p^{+} +\bar\p^{-} +V^{+} +V^{-} )\cdot \nabla\varphi
 + \Delta[(a^{+}-a^{-})V^{-}] ,
 \end{array}
 \end{equation}
note here the last term on the left hand side is a good term, which
ensures the closure of energy estimate for $K$. Note that the lack
of such term in equation of $V^{\pm}$ is the main difficulty in
energy estimate thus we will use the frequency decomposition method
introduced in \cite{lwy09}.

To proceed, we first give the  a priori assumption
\begin{equation}\label{l19}
{\mathcal M}(t)=   \quad \quad\quad\quad\quad \quad \quad\quad\quad\quad
 \quad \quad\quad\quad\quad \quad \quad\quad\quad\quad
\end{equation}
$$
\max \Big\{ \sup\limits_{0\leq s\leq t, |\a|\leq k-2,
p\geq 2}(1+s)^{\frac{n}{2}(1-\frac{1}{p})+\frac{|\a|+1}{2}}
\|\partial^\a_xV^{\pm} (\cdot,s)  \|_{L^p},
\sup\limits_{0\leq
s\leq t, |\a|=k,
k-1}(1+s)^{\frac{n}{4}+\frac{|\a|+1}{2}} \|\partial^\a_xV^{\pm}  (\cdot,s)\|_{L^2},
$$
$$
\sup\limits_{0\leq s\leq t, |\a|\leq
k-2, p\geq 2}(1+s)^{\frac{n}{2}(1-\frac{1}{p})+\frac{|\a|+2}{2}} \|\partial^\a_xU^{\pm} (\cdot,s)\|_{L^p} ,
 \sup\limits_{0\leq s\leq t, |\a|=k,
k-1}(1+s)^{\frac{n}{4}+\frac{k+1}{2}} \|\partial^\a_xU^{\pm}
(\cdot,s)\|_{L^2}\Big\}.
$$
%\end{array}
Under the assumption in Lemma 2.3 and the above a priori assumption,
it is easy to check that for any multi-indies  $\a$ and $\gamma$,
the nonlinear terms in $F^{\pm}$ satisfy
\begin{equation}\label{l20}
 \left\{
\begin{array}{lcr}
\|\P^\a_y F^i \|_{L^p(\R^n_y)} \leq CE_\rho e^{-s},~ ~  |\a|\leq k ,\vspace{2.5mm}\\
\|\P^{\r}_yF^{ij} \|_{L^p(\R^n_y)} \leq C{\mathcal M}^2
(1+s)^{-(n+1+\frac{|\r|}{2})+\frac{n}{2p}}, ~ ~ |\r|\leq k-2.
\end{array}
\right.
\end{equation}
Here we should indicate that the above estimates hold for $p\geq1$.

Now we perform energy estimates. Multiply equation (\ref{l17}) by
$K$ and integrate the resultant equation over $\R^n$, we have
\begin{eqnarray*}
&&\frac{d}{dt}\int_{\R^n}(\frac12K^2+K K_{t})dx- \int_{\R^n}
\Delta(a^{+}K)Kdx + \int_{\R^n}(\bar\p^{+} +\bar\p^{-} +V^{+} +V^{-}
)K^2dx-\int_{\R^n}K^2_tdx\\
&=&\int_{\R^n} K( F^{+}-F^{-} )dx - \int_{\R^n}\nabla (\bar\p^{+}
+\bar\p^{-} +V^{+} +V^{-} )\cdot \nabla\varphi Kdx
 + \int_{\R^n} \Delta[(a^{+}-a^{-})V^{-}] Kdx.
\end{eqnarray*}
Note
$$
 \int_{\R^n} - \Delta(a^{+}K) K dx= \int_{\R^n} \nabla (a^{+}K) \cdot \nabla
 Kdx
 = \int_{\R^n} a^{+} |\nabla K|^{2} dx - {1\over2} \int_{\R^n} (\Delta
 a^{+})K^{2}dx.
$$
Moreover, using Cauchy-Schwarz's and Young's inequality, we have
\begin{eqnarray*}
\int_{{\R}^n} \nabla (\bar\p^{+} +\bar\p^{-} +V^{+} +V^{-} )\cdot
\nabla\varphi  K dx &\leq& \| \nabla\varphi  \|_{L^{\infty}}   \|
\nabla (\bar\p^{+} +\bar\p^{-} +V^{+} +V^{-} ) \|_{L^{2}}   \| K
\|_{L^{2}}\\
&\leq&  C \| \nabla (\bar\p^{+} +\bar\p^{-} +V^{+} +V^{-} )
\|_{L^{2}} \  \| K \|_{H^{1}}^{2} \\
&\leq&   C(E_{\p}+ {\mathcal M}(t)) (1+t)^{-{5\over 4} } \  \| K
\|_{H^{1}}^{2} ,\\
 \int_{\R^n} \Delta[(a^{+}-a^{-})V^{-}]  Kdx   &= &
  \int_{\R^n} \nabla[(a^{+}-a^{-})V^{-}] \cdot \nabla Kdx \\
   &\leq &C\| \nabla[(a^{+}-a^{-})V^{-}]\|_{L^{2}}    \| \nabla K  \|_{L^{2}}\\
  &  \leq &CE_\p e^{-\beta t} {\mathcal
M}(t)((1+t)^{-\frac n4-\frac12} +(1+t)^{-\frac n4-1})  \| \nabla K
\|_{L^{2}},
\end{eqnarray*}
since the Sobolev norm of $a^{+}-a^{-}$ has same exponential decay
as $\bar\p^{+} - \bar\p^{-}$ in Lemma 2.3, and further
\begin{eqnarray*}
\int_{{\R}^n} K F^{\pm}dx \leq \varepsilon_{0} \int_{{\R}^n} K^{2} +
C(\varepsilon_{0}) \int_{{\R}^n} (F^{\pm})^{2} \leq \varepsilon_{0}
\int_{{\R}^n} K^{2} + C(\varepsilon_{0})( E_{\p}^{2}+ {\mathcal
M}(t)^{4}) (1+t)^{-{5\over 2}n-4},
\end{eqnarray*}
here we have used (\ref{l20}) in the above estimate. Combine above
estimates and good decay properties of $\bar\p^{\pm}$ in Lemma 2.3,
we have
\begin{equation}\label{l23}
 \begin{array}{rl}
&{\dis {d \over dt} \int_{{\R}^n} ({1 \over 2} K^{2} + K_{t}K)dx-\int_{{\R}^n} K_{t}^{2} + \int_{{\R}^n}( K^{2}dx +  |\nabla K|^{2})dx  }\vspace{2mm} \\
 \leq &C( E_{\p}^{2}+ {\mathcal M}(t)^{4}) (1+t)^{-{5\over 2}n-4}.
 \end{array}
 \end{equation}
Here and in the subsequent we use the fact that
$\bar\p^{\pm}+V^{\pm}$ is strictly positive and bounded from below.

Next, multiply equation (\ref{l17}) by $K_{t}$ and integrate the
resultant equality over $\R^n$, we have
\begin{eqnarray*}
&&{1\over 2}{d\over dt}\int_{{\R}^n} K_{t}^{2}dx  - \int_{{\R}^n}
\Delta(a^{+}K)K_{t}dx + \int_{{\R}^n} K_{t}^{2}dx  + \int_{{\R}^n}
(\bar\p^{+} +\bar\p^{-} +V^{+} +V^{-} )K K_{t}dx\\
&= &\int_{{\R}^n} K_{t}(F^{+}-F^{-})dx - \int_{{\R}^n} \nabla
(\bar\p^{+} +\bar\p^{-} +V^{+} +V^{-} )\cdot \nabla\varphi K_{t}dx
 + \int_{{\R}^n} \Delta[(a^{+}-a^{-})V^{-}] K_{t}dx.
\end{eqnarray*}
It is easy to compute
\begin{eqnarray*}
&&\int_{{\R}^n} - \Delta(a^{+}K) K_{t}dx = \int_{{\R}^n} \nabla
(a^{+}K) \cdot \nabla K_{t}dx\\
 &= &\int_{{\R}^n} a^{+}  {1\over2} \P _{t}|\nabla K|^{2}dx
 + \int_{{\R}^n} \nabla a^{+}  \cdot ( K \nabla K_{t})dx\\
&=&\int_{{\R}^n} a^{+}  {1\over2} \P _{t}|\nabla K|^{2}dx
 + \int_{{\R}^n} \nabla a^{+}  \cdot  \P_{t}( K \nabla K)dx
 - \int_{{\R}^n} \nabla a^{+}  \cdot  ( K_{t} \nabla K)dx\\
&=&  {d \over dt} \int_{{\R}^n} a^{+}  {1\over2}  |\nabla K|^{2}dx
  - \int_{{\R}^n}\P_{t}a^{+}  {1\over2}  |\nabla K|^{2}dx
  -  {d \over dt} \int_{{\R}^n}{1\over2} \Delta a^{+}  K^{2}dx \\
  & &
   + \int_{{\R}^n}{1\over2} \P_{t}\Delta a^{+}  K^{2}dx
   - \int_{{\R}^n} \nabla a^{+}  \cdot  ( K_{t} \nabla K)dx,
\end{eqnarray*}
in which the last term on the right satisfies
\begin{eqnarray*}
|\int_{{\R}^n} \nabla a^{+}  \cdot  ( K_{t} \nabla K)dx| & \leq  &
C\| \nabla a^{+}     \|_{L^{\infty}} \| K_{t}  \|_{L^{2}}   \| \nabla K  \|_{L^{2}}\\
 & \leq &  \varepsilon_{1} \|    K_{t}     \|_{L^{2}}  +
  C(\varepsilon_{1})  E_{\p}^{2} (1+t) ^{-{3\over 2}}  \| \nabla K  \|_{L^{2}}^{2}.
\end{eqnarray*}
Next, note also that $\bar\p^{\pm}+V^{\pm}$ is strictly positive and bounded
\begin{eqnarray*}
&&\int_{{\R}^n} (\bar\p^{+} +\bar\p^{-} +V^{+} +V^{-} )K K_{t}dx \\
&= &{d \over dt} \int_{{\R}^n}  {1\over2} (\bar\p^{+} +\bar\p^{-} +V^{+} +V^{-} ) K^{2}dx
  - \int_{{\R}^n}  {1\over2} \P_{t}(\bar\p^{+} +\bar\p^{-} +V^{+} +V^{-} ) K^{2} dx.
\end{eqnarray*}
Also note
\begin{eqnarray*}
\int_{{\R}^n} \nabla (\bar\p^{+} +\bar\p^{-} +V^{+} +V^{-} )\cdot
\nabla\varphi  \ K_{t} &\leq& C\| \nabla\varphi  \|_{L^{\infty}}
\| \nabla (\bar\p^{+} +\bar\p^{-} +V^{+} +V^{-} ) \|_{L^{2}}   \|
K_{t}
\|_{L^{2}}\\
&\leq& \varepsilon_{2}  \| K_{t} \|_{L^{2}}^{2} +C(\varepsilon_{2})
( E_{\p}^{2}+ \mathcal M^{2}) (1+t) ^{-{3\over 2}}   \| K
\|_{L^{2}},\\
 \int_{{\R}^n} \Delta[(a^{+}-a^{-})V^{-}] K_{t}dx &\leq&
  \varepsilon_{3}  \| K_{t} \|_{L^{2}}^{2} +C(\varepsilon_{3})
  ( E_{\p}^{2}+ \mathcal M^{2}) t^{-\beta t},
\end{eqnarray*}
and
\begin{eqnarray*}
\int_{{\R}^n} K_{t} F^{\pm} dx&\leq &\varepsilon_{0} \int K_{t}^{2}dx
+ C(\varepsilon_{0}) \int_{{\R}^n} (F^{\pm})^{2}dx \\
&\leq&
\varepsilon_{0} \int_{{\R}^n} K_{t}^{2} dx+ C(\varepsilon_{0})(
E_{\p}^{2}+ {\mathcal M}^{4}) (1+t)^{-{5\over 2}n-4},
\end{eqnarray*}
then combine above estimates to get
\begin{equation}\label{l24}
\begin{array}{rl}
&{\dis {d \over dt} \int_{{\R}^n} ( K_{t}^{2} + K^{2} +|\nabla K|^{2})dx
   +\int_{{\R}^n} K_{t}^{2}dx}  \vspace{2mm}\\
   \leq & {\dis CE_{\p}   \int_{{\R}^n} (K^{2} + |\nabla K|^{2} )dx
   + C( E_{\p}^{2}+ {\mathcal M}^{4}) (1+t)^{-{5\over 2}n-4}.}
   \end{array}
 \end{equation}
Then multiply (\ref{l23}) by ${1\over 4}$  then add to (\ref{l24}),
and if we assume that $ E_{\p} +  \mathcal M$ is small enough, then
$$
 {d \over dt} \int_{{\R}^n}(K^{2} + |\nabla K|^{2}  + K_{t}^{2})dx
 +  \int_{{\R}^n}(K^{2} + |\nabla K|^{2}  + K_{t}^{2})
 \leq C ( E_{\p}^{2}+ {\mathcal M}^{4}) (1+t)^{-{5\over 2}n-4},
$$
by Gronwall's inequality, we have
$$
  \int_{{\R}^n}K^{2} + |\nabla K|^{2}  + K_{t}^{2}
  \leq C  ( E_{\p}^{2}+ {\mathcal M}^{4}) (1+t)^{-{5\over 2}n-4}.
$$
then
$$
  \|K\|_{L^{2}} \leq C  ( E_{\p} + {\mathcal M}^{2}) (1+t)^{-{5\over 4}n-2}.
$$
For higher order derivatives (see also the estimate on high
frequency part of $V^{\pm}$ below), take $\P^{\gamma}$ on both side
of equation \eqref{l17}, multiply by $\P^{\gamma}K$ and
$\P^{\gamma}K_{t}$ and integrate, respectively, then combine as
above estimates for lower oder derivative. Note that the last term
on the right hand side of \eqref{l17} $ \Delta[(a^{+}-a^{-})V^{-}] $
has already second order derivative and the a priori assumption
\eqref{l19} has control of derivatives up to $k$-th order, so we can
only carry out the computation for derivatives with order
$|\gamma|\leq k-2$. Similar arguments as above,  we have,  for
$|\gamma|\leq k-2$,
\begin{equation}\label{l25}
   \| \P^{\gamma}K   \|_{L^{2}} \leq C  ( E_{\p} + {\mathcal M}^{2}) (1+t)^{-{5\over 4}n-2- {|\gamma|\over 2}} ,
\end{equation}
and for $|\gamma|= k-1$,
\begin{equation}\label{sp1}
 \|\P^\gamma_xK \|_{L^2} \leq C \epsilon_0 (1+t)^{-{5\over 4}n-1- {k\over 2}} .
\end{equation}
Combine the estimates (\ref{l21})-(\ref{l22}),
(\ref{l25})-(\ref{sp1}) and using Lemma 3.1, we have the results of
(iii) in Theorem 1.1.

\section{Approximate
Green Function and $L^p$ Estimates on the Low Frequency Component.}
In this section, we will give approximation Green function of the
equation to $V^{\pm}$ as in \cite{lwy09}, which is used to get $L^p$
estimates on the low frequency component of $V^{\pm}$. Recall the
linearized equation (\ref{l15}),
\begin{equation}\label{1027-1}
V^{\pm}_{tt}-\triangle(a^{\pm}(x,t)V^{\pm})+  V^{\pm}_{t}= F^{\pm} \mp \mathrm{div}[( \bar{\p}^{\pm}+V^{\pm} )\nabla\varphi].
\end{equation}
We slightly abuse notations by dropping  `$\pm$' sign without
confusion in the following estimates. Note the main difference of
\eqref{1027-1} to the linearized equation in \cite{lwy09} is that we
need to consider the coupling term `$\mp \mathrm{div}[(
\bar{\p}^{\pm}+V^{\pm} )\nabla\varphi]$' in the present setting.

\subsection{ Approximate Green Function.}
In this subsection, we study the approximate Green function for
(\ref{1027-1}). For convenience of readers, we briefly repeat the
construction of the approximate Green function in below.

Let $G(x,t; y, s)$ be the approximate Green function for the homogeneous part of  (\ref{1027-1}) which meets the basic requirement
$$
G(x,t; y, t)=0,\ \ G_t(x,t; y, t)=\delta (x-y),
$$
where $\delta $ is the Dirac function. Multiplying  (\ref{1027-1}) whose
variables are now changed to $(y, s)$ by $G$ and integrating with
respect to $y$ and $s$ over the region ${\R}^n\times [0, t]$ to get (note here and below we dropped the notation `$\pm$')
\begin{equation}\label{l27}
 \begin{array}{rl}
V(x,t) =&\dis{\int_{{\R}^n}G_s(x,t; y, 0)V(y, 0)dy
}-\dis{\int_{{\R}^n}G(x,t; y, 0)(
V+V_s)(y, 0)dy}\vspace{2.5mm}\\
&+\dis{\int^t_0\int_{{\R}^n} (G_{ss}-a^{+}\triangle_y G-
G_s )(x,t; y, s)V(y, s)dy ds} \vspace{2.5mm}\\
&-\dis{\int^t_0\int_{{\R}^n}G(x,t; y, s)
F(y, s) dy  ds}\vspace{2.5mm}\\
&+\dis{\int^t_0\int_{{\R}^n}G(x,t; y, s) \mathrm{div}  [ (\bar\p
 + V ) \nabla \varphi](y, s) dy ds}.
\end{array}
\end{equation}
If $a (y, s)$ is a constant and $G$ is the Green function of the
homogeneous part of (\ref{1027-1}), then  the third integral in above
is zero. However, when $a (y, s)$ is not a constant, it is
difficult to give an explicit expression of the Green function.
Therefore, we will use the approximate Green function constructed in
\cite{[W-Y3],lwy09}. The idea is to first consider the linear
partial differential equation
\begin{equation}\label{v5-1}
\P_{tt}V-\mu\triangle V+ V_t=0,
\end{equation}
where $\mu$ is a bounded parameter with $C_0<\mu<C_1$, and denote its Green function by $G^\sharp(\mu; x,t)$, whose Fourier transform
$$
\hat{G}^\sharp(\mu;\xi,t)=\frac{e^{\l_+t}-e^{\l_-t}}{\l_+-\l_-},
$$
where
$$
\l_\pm(\xi)\equiv \frac{1}{2}(-1\pm \sqrt{1-4\mu |\xi|^2}).
$$
Denote $\hat{G}^\sharp= \hat{E}^+ + \hat{E}^-$
with
$$
\hat{E}^+=\eta_0e^{\l_+t}, ~~ \hat{E}^-=\eta_0e^{\l_-t},
~~\eta_0=(\l_+ - \l_-)^{-1}.
$$

The approximate Green function is defined by
\begin{equation}\label{p6}
 G(x,t; y, s)=G^\sharp(a (y, \sigma(t,s)); x-y, t-s),
\end{equation}
with $a (y, \sigma(t,s))=P^\prime(\bar{\rho} (y, \sigma(t,s)))$, and the
function $\sigma(t,s)$ is chosen such that $\sigma(t, s)\in C^3([2,
\infty]\times [0, \infty])$,
$$
\sigma(t, s)=\left\{
\begin{array}{rl}
& s, \quad s>t/2,\vspace{2.5mm}\\
&t/2, \quad s\leq t/2-1,
\end{array}\right.
$$
and
$$
\sum_{1\leq l_1+l_2\leq 3} |\P^{l_1}_t\P^{l_2}_s\sigma(t,
s) |\leq C, ~~s\in (t/2-1, t/2).
$$
Notice that $\sigma^{-1}(t, s)\leq C(1+t)^{-1}$ for $t>2$ so that we
have by Lemma 2.1
\begin{equation}\label{p7}
 (1+t) |\P_s a (y, \sigma(t,s)) |+(1+t)^2 |\P^2_s a (y,
\sigma(t,s)) |\leq CE_\rho,
\end{equation}
where $E_\rho$ is defined in Lemma 2.1.

Notice that the decay of the derivatives of the function
$a (y,\sigma(t,s))$ with respect to time will be used in the
following analysis. Recall that the approximate Green function
defined in (\ref{p6}) is not symmetric with respect to the variables
$(x,t)$ and $(y,s)$. However, straightforward calculation gives
their relations as
\begin{equation}\label{p9}
\P_{x_i}G=-\P_{y_i}G + \P_a(G^\sharp)~a_{x_i}, ~~\P_t G=-\P_s G +
\P_a(G^\sharp)~(a_s+a_t).
\end{equation}

Denote the low frequency component in  the approximate Green function
$G(x,t;y,s)$ by
$$%\begin{equation}\label{p8}
 G_L(x,t; y, s)=\chi(D_x)G(x,t; y,s),
$$%\end{equation}
where $\chi(D_x)$, $D_x=\frac{1}{\sqrt{-1}}\P_x
=\frac{1}{\sqrt{-1}}(\P_{x_1},\cdots, \P_{x_n}) $, is the
pseudo-differential operator with symbol $\chi(\xi)$ as a smooth
cut-off function satisfying
$$
\chi(\xi)=\left\{\begin{array}{ll}1,&|\xi|<\varepsilon,\vspace{2.5mm}\\0,&|\xi|>2\varepsilon,
\end{array}\right.%\eqno{(3.12)}
$$
for some chosen constant $\varepsilon$ in $(0, \varepsilon_0)$ with
$\varepsilon_0=\frac{1}{2}\min\Big\{1, \sqrt{\frac{1}{4C_1}}\Big\}$, $C_{1}$
is the upper bound of $\mu$ in \eqref{v5-1}. Moreover, we have
$$
\begin{array}{rl}
G_L(x,t;y,s)=&\frac{1}{(2\pi)^{ n}}\dis{\int_{\R^n}} \chi(\xi)
e^{\sqrt{-1}(x-y)\xi} \hat{G}^\sharp(a(y, \sigma(t,s)),\xi,t-s)
d\xi\vspace{2.5mm}\\
=&G^\sharp_L(a(y, \sigma(t,s)); x-y, t-s).
\end{array}
$$

Therefore, it is direct to have the following proposition
\cite{lwy09}.

\begin{prop}
For $q\in [1, \infty]$ and any indices $h, l, \alpha$ and $\beta$,
we have
$$
\begin{array}{rl}
\displaystyle{\sup_{y}} \|\P^\a_x \P^\b_y \P^l_s \P^h_t
G_L(\cdot,t; y,s) \|_{L^q(\R^n_x)} \leq
C(1+t-s)^{-\frac{n}{2}(1-\frac{1}{q})-\frac{2\min(l+h,1)+|\a|+|\b|}{2}},\vspace{2.5mm}\\
\displaystyle{\sup_{x}} \|\P^\a_x \P^\b_y \P^l_s \P^h_t G_L(x,t;
\cdot,s) \|_{L^q(\R^n_y)} \leq
C(1+t-s)^{-\frac{n}{2}(1-\frac{1}{q})-\frac{2\min(l+h,1)+|\a|+|\b|}{2}}.
\end{array}
$$
\end{prop}

\subsection{$L^p$ Estimates on the Low
Frequency Component.} In this subsection, we will establish the
$L^p$ estimates on the low frequency component by using the
approximate Green function. Assume that $|\a|\leq k$ in this
section. To derive the $L^p$ estimates for the low frequency part,
recall (\ref{l27}), and set
\begin{eqnarray*}
&&I^{\a}_1= \chi(D_x)\int_{{\R}^n}\P^\a_{x}G_s(x, t; y, 0)V(y,0)dy
 = \dis{\int_{{\R}^n}\P^\a_{x}(G_L)_s(x, t; y,0)V(y,0)dy},\\
&&I^\a_2=-\dis{\int_{{\R}^n}\P^\a_{x}G_L(x, t; y, 0)(V+V_s)(y,
0)dy},\\
&&I^{\a}_3= \int^t_0\int_{{\R}^n}\P^\a_{x}R_{G_L}(x, t; y, s)~ V(y,
s)dyds,\\
&&I^{\a}_4= - \dis{\int^t_0\int_{{\R}^n}\P^\a_{x} G_L(x, t; y,
s) F (y, s)dy ds},\\
&&I^{\a}_5 = \dis{\int^t_0\int_{{\R}^n}\P^\a_{x}G_{L}(x,t; y, s)
\mathrm{div} [ (\bar\p  + V ) \nabla
\varphi](y, s) dy  ds} ,
\end{eqnarray*}
where
$$
R_G\equiv G_{ss}(x, t; y, s)-a^{+}(y, s)\triangle_y G(x, t; y, s)-
G_{s}(x, t; y, s),
$$
and
$$
\chi(D_x)R_G=R_{G_L}.
$$
Since
$$
(G^\sharp_{tt}-a \triangle G^\sharp+G^\sharp_t)(a(y, s); x-y,
t-s)=0,
$$
we have
$$
\begin{array}{rl}
&R_{G_L}(x, t; y, s)\\
=& \Big[G^\sharp_{L;0, 0}(a(y, s); x-y, t-s)a_s(y,
\sigma)^2-2G^\sharp_{L;0, n+1}(a(y, s); x-y, t-s)a_s(y, \sigma)\\
&+G^\sharp_{L;0}(a(y, s); x-y, t-s) a_{ss}(y, \sigma) -
G^\sharp_{L;0}(a(y, s); x-y,
t-s) a_s(y, \sigma)\\
&+a(y,s) \Big(\sum\limits_{i=1}^n [G^\sharp_{L;0, i}(a(y, s); x-y,
t-s)a_{y_i}(y,
\sigma)\\
&-G^\sharp_{L;0, 0}(a(y, s); x-y, t-s)((a)^2_{y_i})(y, \sigma)] +
G^\sharp_{L;0}(a(y, s); x-y,
t-s)\triangle_{y}a(y,\sigma)  \Big) \Big]\vspace{2.5mm}\\
& + [(a(y,
\sigma)-a(y, s))
\triangle G^\sharp_{L}(a(y, s); x-y, t-s) ]\vspace{2.5mm}\\
=: &R^1_{G_L}+R^2_{G_L}.
\end{array}
$$
Here $R^i_{G_L}$, $i=1,2$,  is the corresponding term in the above
summation in the above equation. To denote the derivatives, we use
the notations $G^\sharp_{L;0}(a;x,t)=\partial_aG^\sharp_{L}(a;x,t)$,
$G^\sharp_{L;i}(a;x,t)=\partial_{x_i}G^\sharp_{L}(a;x,t)$,
$G^\sharp_{L;n+1}(a;x,t)=\partial_tG^\sharp_{L}(a;x,t)$,
$G^\sharp_{L;0,i}(a;x,t)=\partial_a\partial_{x_i}G^\sharp_{L}(a;x,t)$,
and
$G^\sharp_{L;0,n+1}(a;x,t)=\partial_a\partial_tG^\sharp_{L}(a;x,t)$
etc..

Then, set $X_L(x, t)=\chi(D_x)X(x, t)$, from above notations we
have
$$
\P^\a_{x}X_L(x,t)=I^{\a}_1+I^{\a}_2+I^{\a}_3+I^\a_4 +I^{\a}_5.
$$
We will estimate the right hand side of above term by term. The
terms from $I^{\a}_1$ to $I^{\a}_4$ are similar to the estimates in
\cite{lwy09}: by Proposition 4.1, it is straightforward to obtain
$$
\begin{array}{rl}
 \|I^\a_1 \|_{L^p(\R^n_x)} \leq
&C(1+t)^{-\frac{n}{2}(1-\frac{1}{p})-\frac{|\a|+2}{2}} \|V_0 \|_{L_1}.
\end{array}
$$
For $I^\a_2$, set
$$
\tilde{\nu}_0 (y)=\nu_t (y, 0)+\nu (y, 0),
$$
where $\nu$ (with ${\pm}$ omitted) is defined in (\ref{l6}). Then
$$
\begin{array}{rl}
 |I^\a_2 |= |\dis{\int_{{\R}^n}}\P_{y_1}\P^\a_{x}G_L(x,t;
y, 0) \tilde{\nu}_0  (y)dy |.
\end{array}
$$
Also by using Propositions 4.1, we have
$$
\begin{array}{rl}
 \|I^\a_2 \|_{L^p(\R^n_x)} \leq
C(1+t)^{-\frac{n}{2}(1-\frac{1}{p})-\frac{|\a|+1}{2}}   \|\tilde{\nu}_0  \|_{L_1} .
\end{array}
$$

We now turn to estimate the  term $I^\a_3$ which is the error coming
from the approximate Green function. For illustration,  we only
consider
$$
J^\a_1=\int^t_{0}\int_{{\R}^n}\P^\a_{x}G^\sharp_{L;0}(a^{+}(y,\sigma), x-y, t-s)a^{+}_s(y,\sigma)V(y, s)dyds,
$$
and
$$
J^\a_2=\int^t_{0}\int_{{\R}^n}\P^\a_{x}R^2_{G_L}(x,t; y, s)V(y, s)dy ds,
$$
because the other terms in $I^\a_3$ can be estimated similarly. Note
that (\ref{p7}) gives
$$
|a^{+}_s(y, \sigma)|\leq CE_\rho (1+t)^{-1},
$$
then we have, for $|\r|\leq k-2$,
$$
\begin{array}{rl}
 \|J^\r_1 \|_{L^p(\R^n_x)}\leq &
\dis{\int^t_{0} \|\int_{{\R}^n}}\P^\r_{x}G^\sharp_{L;0}(a(y,\sigma),
x-y, t-s) ~a_s(y,\sigma)~V(y, s)dy \|_{L^p{(\R^n_x)}}ds\vspace{2.5mm}\\
\leq &CE_\rho{\mathcal
M}\Big[\dis{\int^{t/2}_0}(1+t-s)^{-\frac{n}{2}(1-\frac{1}{p})-\frac{|\r|}{2}}(1+t)^{-1} (1+s)^{-\frac{n}{2}(1-\frac{1}{1})-1/2}ds\vspace{2.5mm}\\
&+\dis{\int^t_{t/2}}(1+t-s)^{-\frac{n}{2}(1-\frac{1}{1})} (1+t)^{-1}
(1+s)^{-\frac{n}{2}(1-\frac{1}{p})-\frac{|\r|+1}{2}}ds\Big]\vspace{2.5mm}\\
\leq & CE_\rho{\mathcal
M}(1+t)^{-\frac{n}{2}(1-\frac{1}{p})-\frac{|\r|+1}{2}},
\end{array}
$$
and for $|\r|= k-1$ and $k$,
$$
\begin{array}{rl}
 \|J^\r_1 \|_{L^p(\R^n_x)} \leq &CE_\rho{\mathcal
M}\dis{\int^{t/2}_0}(1+t-s)^{-\frac{n}{2}(1-\frac{1}{p})-\frac{|\r|}{2}}
(1+t)^{-1} (1+s)^{-\frac{n}{2}(1-\frac{1}{1})-1/2}ds\vspace{2.5mm}\\
&+CE_\rho{\mathcal M}\dis{\int^t_{t/2}}(1+t-s)^{-\frac{|\r|+2-k}{2}}
(1+t)^{-1}
(1+s)^{-\frac{n}{2}(1-\frac{1}{p})-\frac{k-2+1}{2}}ds\vspace{2.5mm}\\
\leq & CE_\rho{\mathcal
M}(1+t)^{-\frac{n}{2}(1-\frac{1}{p})-\frac{|\r|+1}{2}}.
\end{array}
$$
For $J^\a_2$, since
$$
\begin{array}{rl}
|a (y, s)-a (y,\sigma)|\leq\int^\sigma_s |a_\tau (y, \tau)|d\tau
\leq\left\{
\begin{array}{rl}
&CE_\rho\Theta(t, s), s<t/2,\vspace{2.5mm}\\
&0, \quad s\geq t/2,
\end{array}\right.
\end{array}
$$
where
$$
\Theta(t, s)=(1+t-s)(1+t)^{-1+1/h}(1+s)^{-1/h},
$$
and $h$ can be any positive integer. By using Proposition 4.1, we
have
$$
 \|J^\a_2 \|_{L^p(\R^n_x)} \leq CE_\rho{\mathcal M}
\dis{\int^{t/2}_0}
(1+t-s)^{-\frac{n}{2}(1-\frac{1}{p})-\frac{|\a|+2}{2}}\Theta(t,
s)(1+s)^{-1/2}ds.
$$
By noticing that
$$
\begin{array}{rl}
\dis{\int^{t/2}_0}(1+t)^{-1+1/h}(1+s)^{-1/h}(1+s)^{-1/2}ds
=&(1+t)^{-1+1/h}(1+s)^{\frac{1}{2}-\frac{1}{h}} \Big|^{t/2}_0\vspace{2.5mm}\\
\leq & C(1+t)^{-1+1/h}(1+t)^{\frac{1}{2}-\frac{1}{h}}\ =\ C(1+t)^{-\frac{1}{2}},
\end{array}
$$
we obtain
$$
 \|J^\a_2 \|_{L^p(\R^n_x)}\leq CE_\rho{\mathcal M}
(1+t)^{-\frac{n}{2}(1-\frac{1}{p})-\frac{|\a|+1}{2}}.
$$
Thus, combine the above estimate to have
$$
 \|I^\a_3 \|_{L^p(\R^n_x)}\leq CE_\rho{\mathcal M}
(1+t)^{-\frac{n}{2}(1-\frac{1}{p})-\frac{|\a|+1}{2}}.
$$

Next for $I^\a_4$, recall that $F$ (with $\pm$ omitted) satisfies (\ref{l20}) under
the a priori assumption (\ref{l19}), then for $|\r|\leq k-2$,
$$
\begin{array}{rl}
 \|I^\r_4 \|_{L^p(\R^n_x)}\leq&\dis{\int^t_0 \|
\int_{{\R}^n}\P^\r_{x}G_L
~F dy \|_{L^p(\R^n_x)}ds}\vspace{2.5mm}\\
=&\dis{\int^{\frac{t}{2}}_0 \| \int_{{\R}^n}(-\sum
\P^\r_{x}\P_{y_i}G_L
~F^i +\sum\P^\r_{x}\P_{y_iy_j}G_L ~ F^{ij})dy \|_{L^p(\R^n_x)}ds}\vspace{2.5mm}\\
&+\dis{\int^t_{\frac{t}{2}} \|\int_{{\R}^n} \P^\r_{x}G_L ~F dy \|_{L^p(\R^n_x)}ds}\vspace{2.5mm}\\
\leq &CE_\rho\dis{\int^{\frac{t}{2}}_0}
(1+t-s)^{-\frac{n}{2}(1-\frac{1}{p})-\frac{|\r|+1}{2}} e^{- s} ds
+CE_\rho\dis{\int^t_{\frac{t}{2}}} (1+t-s)^{-\frac{n}{2}(1-\frac{1}{p})-\frac{|\r|+1}{2}} e^{- s} ds\vspace{2.5mm}\\
 &+C{\mathcal M}^2\dis{\int^\frac{t}{2}_0 }(1+t-s)^{-\frac{n}{2}(1-\frac{1}{p})-\frac{|\r|+2}{2}} (1+s)^{-(n+1)+\frac{n}{2}} ds\vspace{2.5mm}\\
 &+C{\mathcal M}^2\dis{\int^t_{\frac{t}{2}}} (1+t-s)^{-1} (1+s)^{-(n+1+\frac{|\r|}{2})+\frac{n}{2p}} ds\vspace{2.5mm}\\
\leq &C(E_\rho+{\mathcal M}^2)
(1+t)^{-\frac{n}{2}(1-\frac{1}{p})-\frac{|\r|+1}{2}}.
\end{array}
$$
The cases when $|\r|= k-1$ and
$k$ can be estimated similarly, and the only
difference is the estimation on the terms like
$$\int^t_{\frac{t}{2}} \|\int_{\R^n}\P^\r_{x}\P_{y_iy_j}G_L ~
F^{ij}dy \|_{L^p(\R^n_x)}ds,\qquad |\gamma|\le k-2.
$$
On the other hand, these terms can be estimated by replacing the
derivatives of $G_L$ w.r.t. $x$ to the derivatives of $G_L$ w.r.t.
$y$  using (\ref{p9}). Then by using integration by parts $k-2$
times to transfer the derivatives on $G_L$ to $F^{ij}$, we have by
(4.8) and Proposition 4.1 that
$$
\begin{array}{rl}
\dis{\int^t_{\frac{t}{2}} \|\int_{\R^n}}\P^\r_{x}G_L ~\P_{y_iy_j}
F^{ij}dy \|_{L^p(\R^n_x)}ds
\leq& C{\mathcal M}^2\dis{\int^t_{\frac{t}{2}}}(1+t-s)^{-\frac{|\r|+2-(k-2)}{2}}(1+s)^{-(n+1+\frac{k-2}{2})+\frac{n}{2p}}ds\vspace{2.5mm}\\
\leq&C{\mathcal M}^2  (1+t)^{-(n+1+\frac{k-2}{2})+\frac{n}{2p}}\vspace{2.5mm}\\
\leq&C{\mathcal M}^2
(1+t)^{-\frac{n}{2}(1-\frac{1}{p})-\frac{|\r|+1}{2}}.
\end{array}
$$
Therefore, we have the $L^p$ estimate on $I^\a_4$ as
$$
 \|I^\a_4 \|_{L^p(\R^n_x)}\leq C{\mathcal M}^2
(1+t)^{-\frac{n}{2}(1-\frac{1}{p})-\frac{|\a|+1}{2}}.
$$

For $I^{\a}_5$, we write
\begin{eqnarray*}
I^{\a}_5 &=& \int^{t\over 2}_0 \int_{{\R}^n}\P^\a_{x}
\nabla_{y}G_{L}(x,t; y, s)
 \cdot [ (\bar\p + V ) \nabla \varphi](y, s) dy
 ds\\
 &&+ \int^t_{t\over 2}\int_{{\R}^n}  \P^\a_{x}G_{L}(x,t; y, s)
 \mathrm{div}[ (\bar\p  + V ) \nabla \varphi](y, s) dy  ds\\
&=&: I^{\a}_{5,1} + I^{\a}_{5,2},
\end{eqnarray*}
in which the first term satisfies
$$
 I^{\a}_{5,1} \leq  C\int^{t\over 2}_0
 \|   \P^\a_{x} \nabla_{y} G_{L} \|_{L^{P}}
  \|   (\bar\p  + V ) \nabla \varphi  \|_{L^{1}}ds .
$$
To estimate
$$
\|   (\bar\p  + V ) \nabla \varphi \|_{L^{1}} \leq
C\| \bar\p + V \|_{L^{6\over 5}} \|  \nabla \varphi \|_{L^{6}},
$$
we note that the $L^{6\over 5}$ norm of   $\bar\p  + V( = \p, positive)$  can be
controlled by its $ L^{1} $ and $L^{2}$ norms by interpolation, and
also note the $ L^{1} $ norm of $\p$ (with $\pm$ omitted) is conserved because the conservation of mass, $ L^{2} $ norm can be bounded by the a priori assumption, then we have
$$
\| I^{\a}_{5,1} \|_{L^{P}} \leq C{\mathcal M}^2
(1+t)^{-\frac{n}{2}(1-\frac{1}{p})-\frac{|\a|+1}{2}}.
$$
For the term $ I^{\a}_{6,2}$, we estimate
$$
 I^{\a}_{5,2} \leq C \int^t_{t\over 2}
  \|   \P^\a_{x}   G_{L} \|_{L^{1}}
  \| \mathrm{div} [ (\bar\p  + V ) \nabla \varphi]  \|_{L^{p}}ds .
$$
Note that both $\mathrm{div} \nabla \varphi  =K$ and $ \nabla
\varphi$ are in $ L^{\infty}$ thus have good decay properties, then the above
term decays faster than that of $ I^{\a}_{5,1} $, then we have
$$
\|  I^{\a}_{5} \|_{L^{P}} \leq
C{\mathcal M}^2
(1+t)^{-\frac{n}{2}(1-\frac{1}{p})-\frac{|\a|+1}{2}}.
$$

In summary, by combining all above estimates, we have the estimates
on the low frequency component of $ V^\pm$ in the following theorem.
\vspace{2mm}

{\large T}{\footnotesize HEOREM} 4.2. {\it For $|\a|\leq k$, we
have,
$$
 \|\P^\a  V^{\pm}_L   (t) \|_{L^p} \leq C(E_0+{\mathcal M}^2)
(1+t)^{-\frac{n}{2}(1-\frac{1}{p})-\frac{|\a|+1}{2}},
$$
where $E_0=\max \{  \|V^{\pm}_0 \|_{L^1},
 \|\tilde{\nu}^{\pm}_0 \|_{L^1},  \|(V^{\pm}_0, U^{\pm}_0) \|_{H^k},
 \|V^{\pm}_t(0) \|_{H^{k-1}}, E_\p  \}$.}

As an immediate consequence, we have  the $L^2$ estimate on  the
derivatives of order higher than the $k$-th for the low frequency
component because
$$
 \|\P_{x_i}\P^\a V^{\pm}_L(t) \|_{L^2}= \|\xi_i\xi^\a
\chi(\xi)\hat{V^{\pm}}  \|_{L^2} \leq \varepsilon  \|\xi^\a
\chi(\xi)\hat{V^{\pm}} \|_{L^2}=\varepsilon  \|\P^\a
V^{\pm}_L(t) \|_{L^2}.
$$
Thus, we have the following corollary.

\begin{coro}
For any $|\r|> k$, we have
$$
 \|\P^\r V^{\pm}_L(t) \|_{L^2}\leq  C(E_0+{\mathcal M}^2)\varepsilon
(1+t)^{-\frac{n}{2}(1-\frac{1}{2})-\frac{k+1}{2}}.
$$
\end{coro}

\section{Estimates
on the High Frequency Component.}
In this section, we will carry out
the energy estimates on the high frequency component. Recall the
linearized equation (\ref{l15}), set $\tilde{\chi}(\xi)=1-\chi(\xi)$
and $V^{\pm}_H(x, t)=\tilde{\chi}(D_x)V^{\pm}(x, t)$. By taking
$\tilde{\chi}(D_x)$ on both sides of (\ref{l15}) and integrating its
product with $V^{\pm}_H$ and $(V^{\pm}_H)_t$ over $\R^n$
respectively, we have
\begin{equation}\label{p1}
 \frac{d}{dt}\int_{{\R}^n}V^{\pm}_H~(V^{\pm}_H)_t dx -
\int_{{\R}^n}((V^{\pm}_H)_t)^2 dx -\int_{{\R}^n}V^{\pm}_H ~
\triangle\tilde{\chi}(a V^{\pm})dx
+\frac{d}{dt}\int_{{\R}^n}\frac{1}{2}(V^{\pm}_H)^2 dx
\end{equation}
$$
=\int_{{\R}^n}V^{\pm}_H\tilde{\chi} F^{\pm} dx
 \mp \int_{{\R}^n}(V^{\pm}_H)_{t}\tilde{\chi}
 \mathrm{div}[( \bar{\p}^{\pm}+V^{\pm} )\nabla\varphi] ,
$$
and
\begin{equation}\label{p2}
 \frac{d}{dt}\int_{{\R}^n}\frac{1}{2}((V^{\pm}_H)_t)^2 dx  -\int_{{\R}^n}(V^{\pm}_H)_t ~\triangle\tilde{\chi}(a V^{\pm})dx
+\int_{{\R}^n}((V^{\pm}_H)_t)^2dx
\end{equation}
$$=\dis{\int_{{\R}^n}(V^{\pm}_H)_t\tilde{\chi}F^{\pm} dx}
  \mp  \int_{{\R}^n}V^{\pm}_H\tilde{\chi} \mathrm{div}[( \bar{\p}^{\pm}+V^{\pm} )\nabla\varphi].
$$
Again, we slightly abuse notations by dropping the $\pm$ sign without
confusion in the following estimates. First for the third term on
the left hand side of (\ref{p1}) as follows. That is,
$$
\begin{array}{rl}
\dis{-\int_{{\R}^n}V_H ~\triangle\tilde{\chi}(a V)dx }
=\dis{\int_{{\R}^n}a|\nabla V_H|^2dx-\int_{{\R}^n}V_H~\nabla [\nabla
\tilde{\chi}, a]Vdx,}
\end{array}
$$
where $[A, B]=A\circ B-B\circ A$ denotes the commutator. Since
$$
(1+t)^{1/2} \|\nabla a \|_{L^\infty}+(1+t) \|\triangle
a \|_{L^\infty}\leq CE_\rho,
$$
where $E_\rho$ is defined in Theorem 2.1. It is straightforward  to
show that
$$
\int_{{\R}^n} |\nabla [\nabla \tilde{\chi}, a]V |^2dx\leq
CE^2_\rho{\mathcal M}^2 (1+t)^{-\frac{n}{2}-3},
$$
where ${\mathcal M}$ is defined in (3.4). Thus
$$
 |\int_{{\R}^n}V_H~\nabla [\nabla \tilde{\chi}, a]Vdx  |
\leq \eta\int_{{\R}^n}| V_H|^2dx+ CE^2_\rho {\mathcal
M}^2(1+t)^{-\frac{n}{2}-3}.
%\end{array}\eqno{(5.5)}
$$
We now turn to estimate the second term on the left hand side in
(\ref{p2}). That is,
$$
\begin{array}{rl}
-\int_{{\R}^n}(V_H)_t ~\triangle\tilde{\chi}(a V)dx
=&\dis{\int_{{\R}^n}(\nabla V_H)_t ~a\nabla\tilde{\chi}(
V_H)dx+\int_{{\R}^n}(\nabla V_H)_t
[\nabla\tilde{\chi}, a] V dx}\vspace{2.5mm}\\
=&\dis{\frac{1}{2}\frac{d}{dt}\int_{{\R}^n}a|V_H|^2dx-\frac{1}{2}\int_{{\R}^n}a_t|V_H|^2dx-\int_{{\R}^n}(
V_H)_t \nabla[\nabla\tilde{\chi}, a] V dx,}
\end{array}
$$
in which, we have
$$
|\int_{{\R}^n}a_t |\nabla(V_H) |^2 dx  | \leq
CE^2_\rho(1+t)^{-2}\int_{{\R}^n} |\nabla(V_H) |^2 dx,$$ and
$$|\int_{{\R}^n} (V_H)_t\nabla[\nabla \tilde{\chi}, a]Vdx | \leq
\eta\int_{{\R}^n} |(V_H)_t |^2dx+ CE^2_\rho {\mathcal
M}^2(1+t)^{-\frac{n}{2}-3}.
$$

For $\int_{{\R}^n}V_H\tilde{\chi}F dx$ and
$\int_{{\R}^n}(V_H)_t\tilde{\chi}F dx$ on the right hand side of
(\ref{p1}) and (\ref{p2}), by using Lemma 2.3 and the definition of
${\mathcal M}$ in (\ref{l19}), we have
$$
\begin{array}{rl}
 \|F^i \|_{L^2} \leq &CE_\rho  e^{- s},\vspace{2.5mm}\\
 \|\P^{\r}F^{ij} \|_{L^2} \leq &C{\mathcal M}^2
(1+t)^{-(n+1+\frac{|\r|}{2})+\frac{n}{4}},~~ |\r|\leq k-2,
\end{array}
$$
where $F$ and $F^i, F^{ij}$ defined in (\ref{l18}). Further, it is
straightforward to check that
$$
 |\int_{{\R}^n}V_H\tilde{\chi}F dx |\leq
\eta\int_{{\R}^n} |V_H |^2dx + C(\eta)(E^2_\rho +{\mathcal M}^4)
 (e^{- t}+ (1+t)^{-2(n+1)+\frac{n}{2}}  ),
 $$
and
$$
 |\int_{{\R}^n}(V_H)_t\tilde{\chi}F dx |\leq
\eta\int_{{\R}^n} |(V_H)_t |^2dx +   C(\eta)(E^2_\rho +{\mathcal
M}^4)  (e^{- t}+ (1+t)^{-2(n+1)+\frac{n}{2}}  ).
$$

For the term   $\int_{{\R}^n}(V_H)_{t}\tilde{\chi} \mathrm{div}[(
\bar{\p}+V)\nabla\varphi] $, we have
\begin{eqnarray*}
& & \int_{{\R}^n}(V_H)_{t}\tilde{\chi} \mathrm{div}[(
 \bar{\p}+V)\nabla\varphi]dx \\
 &\leq&
 \varepsilon \int_{{\R}^n} (V_H)_{t}^{2}dx + C(\varepsilon) \int_{{\R}^n} (\mathrm{div}[( \bar{\p}+V)\nabla\varphi]
 )^{2}dx\\
&=&\varepsilon \int_{{\R}^n} (V_H)_{t}^{2}dx +
 C(\varepsilon) \int_{{\R}^n} [\nabla( \bar{\p}+V)\cdot \nabla\varphi
 +(\bar{\p}+V)K
 ]^{2}dx\\
&\leq&  \varepsilon \int_{{\R}^n} (V_H)_{t}^{2}dx +
2 C(\varepsilon) \Big( \| \nabla( \bar{\p}+V)\|^{2}_{L^{2}}
  \|  \nabla\varphi  \|^{2}_{L^{\infty}}
   +  \|\bar{\p}+V\|_{L^{\infty}}^{2}  \|K\|_{L^{2}}^{2}
 \Big)\\
&\leq&  \varepsilon \int_{{\R}^n} (V_H)_{t}^{2} dx+
 C(\varepsilon) (E_{\p}^{2} + \mathcal M^{2}) (1+t) ^{-{3\over 2}n - |\a|-2}.
\end{eqnarray*}
The term $ \int_{{\R}^n}V_H\tilde{\chi} \mathrm{div}[( \bar{\p}+V
)\nabla\varphi]dx $ can be estimated similarly.

To close the energy estimate, one needs the following important fact
about the high frequency part:
$$
\int_{{\R}^n} |\nabla V_H |^2dx\geq
\varepsilon\int_{{\R}^n} |V_H |^2dx.
$$
This is a Poincar\'e type inequality which holds only for the high
frequency part in the whole space. By integrating (\ref{p1})
and (\ref{p2}) over $[0, t]$ and multiplying (\ref{p1}) by some
suitably chosen constant $0<\lambda<1$, when $\eta$ is small, the
combination of above estimates give
$$
\begin{array}{rl}
&\dis{\int_{{\R}^n} (|V_H|^2 + |(V_H)_t|^2 + |\nabla V_H|^2
 )(t)dx
+ \mu\int^t_0\int_{{\R}^n} (|V_H|^2 + |(V_H)_s|^2 + |\nabla V_H|^2 ) dxds }\vspace{2.5mm}\\
\leq&\dis{C \Big[\int_{{\R}^n} (|V_H|^2 + |(V_H)_t|^2 + |\nabla
V_H|^2 )(0)dx+(E^2_\rho+{\mathcal
M}^4)\int^t_0(1+s)^{-\frac{n}{2}-3}ds\Big],}
\end{array}
$$
for some positive $\mu$. Denote
$$
{\mathcal F}(t)  = \int_{{\R}^n} (|V_H|^2 + |(V_H)_t|^2 + |\nabla
V_H|^2 ) dx.
$$
Then the above inequality  gives
$$
{\mathcal F}(t) +\mu\int^t_0{\mathcal F}(s)ds\leq C ({\mathcal
F}(0)+(E^2_\rho+{\mathcal
M}^4)\int^t_0(1+s)^{-\frac{n}{2}-3}ds ).%\eqno{(5.12)}
$$
By using the Gronwall inequality, we have
$$
{\mathcal F}(t) \leq Ce^{-\mu t} ({\mathcal F}(0)+(E^2_\rho+{\mathcal
M}^4)\int^t_0e^{\mu s}(1+s)^{-\frac{n}{2}-3}ds ).
$$
Hence, we have
$$\begin{array}{rl}
& \| V_H(t) \|^2_{H^1} +  \|(V_H)_t(t) \|^2_{L^2} \vspace{2.5mm}\\
 \leq &e^{-\mu t}( \| V_H(0) \|_{H^1} +
 \|(V_H)_t(0) \|_{L^2})\vspace{2.5mm}
 + C(E^2_\rho+{\mathcal M}^4)(1+t)^{-\frac{n}{2}-3}\vspace{2.5mm}\\
 \leq &C(E^2_0+{\mathcal M}^4)(1+t)^{-\frac{n}{2}-3},
 \end{array}    %\eqno{(5.13)}
$$
where $E_0$ is defined in Theorem 4.2.

Next, we will derive the  energy estimates on the higher order
derivatives of  the
high frequency component, that is,  $ \int_{{\R}^n}|\P^\a
V_H|^2 + |\P^\a (V_H)_t|^2 + |\nabla\P^\a V_H|^2 dx $ for $0<
|\a|\leq k-1$. In the rest of this section, we  assume $0<
|\a|\leq k-1$. The estimation can be obtained by
induction on $|\alpha|$.  Assume that
$$ \int_{{\R}^n} (|\P^\r V_H|^2 + |\P^\r
(V_H)_t|^2 + |\nabla\P^\r V_H|^2 ) dx \leq C(E^2_0+{\mathcal
M}^4)(1+t)^{-\frac{n}{2}-(|\r|+3)}  %\eqno{(5.14)}
$$
holds for any multi-index $\r$ with $|\r| < |\a|$, we want to prove
\begin{equation}\label{p13}
 \int_{{\R}^n} (|\P^\a V_H|^2
+ |\P^\a (V_H)_t|^2 + |\nabla\P^\a V_H|^2  )dx \leq C(E^2_0+{\mathcal
M}^4)(1+t)^{-\frac{n}{2}-(|\a|+3)}.
\end{equation}

Taking $\P^\a\tilde{\chi}$ on (\ref{l15}), neglecting the $\pm$ sign without confusing,  and integrating its product with
$\P^\a V_H$ and $\P^\a (V_H)_t$ over $\R^n$ respectively, we have
\begin{equation}\label{p3}
 \frac{d}{dt}\int_{{\R}^n}\P^\a V_H\P^\a (V_H)_t dx -
\int_{{\R}^n}|\P^\a (V_H)_t|^2 dx -\int_{{\R}^n}\P^\a V_H
~\triangle\P^\a \tilde{\chi}(a V)dx
+\frac{d}{dt}\int_{{\R}^n}\frac{1}{2}|\P^\a V_H|^2 dx
\end{equation}
$$
=\int_{{\R}^n}\P^\a V_H~\tilde{\chi}\P^\a F dx \mp
\int_{{\R}^n}\P^\a V_H~\tilde{\chi}\P^\a \mathrm{div}[(
\bar{\p}^{\pm}+V^{\pm} )\nabla\varphi] ,
$$
and
\begin{equation}\label{p4}
 \frac{d}{dt}\int_{{\R}^n}\frac{1}{2}|\P^\a (V_H)_t|^2 dx
-\int_{{\R}^n}\P^\a (V_H)_t ~\triangle\tilde{\chi}\P^\a (a V)dx
+\int_{{\R}^n}|\P^\a (V_H)_t|^2 dx
\end{equation}
$$
=\int_{{\R}^n}\P^\a (V_H)_t~\tilde{\chi}\P^\a F dx\mp
\int_{{\R}^n}\P^\a (V_H)_{t}~\tilde{\chi}\P^\a \mathrm{div}[(
\bar{\p}^{\pm}+V^{\pm} )\nabla\varphi] .
$$
For the third term on the left hand side of (\ref{p3}), we have
$$
%\begin{array}{rl}
\dis{-\int_{{\R}^n}\P^\a V_H ~\triangle\tilde{\chi}\P^\a (a V)dx }%\vspace{2.5mm}\\
=\dis{\int_{{\R}^n}a(\P^\a \nabla V_H)^2dx -\int_{{\R}^n}\P^\a V_H
~\nabla [\nabla\tilde{\chi}\P^\a, a] Vdx.}
$$
Since
$$
 \|\P^\beta_x a \|_{L^\infty}\leq CE_\rho(1+t)^{-|\beta|/2},
$$
it holds that
$$
 |\int_{{\R}^n}\P^\a V_H ~\nabla [\nabla\tilde{\chi}\P^\a, a]
Vdx |\leq \eta\int_{{\R}^n}|\P^\a V_H|^2dx+C_\eta E^2_\rho{\mathcal
M}^2(1+t)^{-\frac{n}{2}-3-|\alpha|}.
$$
Similarly, for the second term on the left hand side of (\ref{p4}), we
have
$$
\dis{-\int_{{\R}^n}\P^\a (V_H)_t ~\triangle\tilde{\chi}\P^\a (a V)dx}
=\dis{\frac{d}{dt}\int_{{\R}^n}\frac{a}{2}|\nabla \P^\a V_H|^2dx
-\int_{{\R}^n}\frac{a_t}{2}|\nabla \P^\a V_H|^2dx}
$$
$$
~\hspace{3cm} \dis{-\int_{{\R}^n}\P^\a(V_H)_t~\nabla[\nabla\P^\alpha\tilde{\chi},
a] Vdx,}
$$
where
$$
 |\int_{{\R}^n}\P^\a (V_H)_t ~\nabla [\nabla\tilde{\chi}\P^\a, a]
Vdx |\leq \eta\int_{{\R}^n}|\P^\a (V_H)_t|^2dx+C_\eta
E^2_\rho{\mathcal M}^2(1+t)^{-\frac{n}{2}-3-|\alpha|}.
$$
For the terms $\int_{{\R}^n}\P^\a V_H\tilde{\chi}\P^\a F dx$ and
$\int_{{\R}^n}\P^\a (V_H)_t\tilde{\chi}\P^\a F dx$ on the right hand
side of (\ref{p3}) and (\ref{p4}), we only estimate the second one because the estimation on the first is easier. Notice that the estimation on the terms
with derivatives of order less or equal to $|\a|+1$ follows directly
from the definition of ${\mathcal M}$ in (\ref{l19}). Thus, we consider the terms with
derivatives of order higher than $|\a|+1$. Firstly, by using the
expression (\ref{l15}) for $F$, we have
\begin{equation}\label{p15}
 \begin{array}{rl}
F=&\tilde{Q}+\triangle ({\mathcal P}_1(\bar{\p}, V)V^2)\vspace{2.5mm}\\
=& [(R_\rho)_t+R_\rho ]-(1
+\partial_t)(V\bar{u}_1)_{x_1}-\mathrm{div}((\bar{\p}+V)_tU)
-\mathrm{div}((\bar{\p}+V)H)+\triangle ({\mathcal P}_1(\bar{\p},
V)V^2).
\end{array}
\end{equation}
Since (\ref{l7}) implies
\begin{equation}\label{p16}
{\rm div}U=-(\bar{\p}+V)^{-1} (V_{t}+(\bar{u}+U)\cdot\nabla V+
(U\cdot\nabla)\bar{\p}+V{\rm div}\bar{u}-R_\p ),
\end{equation}
substituting (\ref{p16}) in (\ref{p15}), by the definition of $H$, we have
$$
F=(\bar{u}+U)\cdot\nabla((\bar{u}+U)\cdot\nabla V) + \triangle
({\mathcal P}_1(\bar{\p}, V)V^2) + {\mathcal R},
$$
where ${\mathcal R}$ denotes the remainder which contains derivatives of
$U$ and $K$ with order at most 1. Thus,  $\P^\a {\mathcal R}$ has
derivatives with order at most $|\a|+1~(\leq k)$. Then
$$
\int_{{\R}^n}\P^\a (V_H)_t ~ \tilde{\chi}\P^\a F dx=N_1 +N_2+N_3,
$$
with
$$
\begin{array}{rl}
N_1=&\dis{\int_{{\R}^n}}\P^\a(V_H)_t
\tilde{\chi}\P^\a ((\bar{u}+U)\cdot\nabla((\bar{u}+U)\cdot\nabla
V) ) dx,\vspace{2.5mm}\\
 N_2=&\dis{\int_{{\R}^n}}\P^\a(V_H)_t
\tilde{\chi}\P^\a\triangle ({\mathcal P}_1(\bar{\p}, V)V^2)dx,\vspace{2.5mm}\\
 N_3=&\dis{\int_{{\R}^n}}\P^\a(V_H)_t
\P^\a{\mathcal R}dx.
\end{array}
$$

For $N_1$, we have
$$
\begin{array}{rl}
N_1=& \dis{\int_{{\R}^n}}\P^\a(V_H)_t
\tilde{\chi}(\bar{u}+U)\cdot\nabla((\bar{u}+U)\cdot\nabla\P^\a V)
dx+\{\cdots\}\vspace{2.5mm}\\
=&\dis{\int_{{\R}^n}}\P^\a(V_H)_t
~(\bar{u}+U)\cdot\nabla((\bar{u}+U)\cdot\nabla\P^\a V_H)
dx+\dis{\int_{{\R}^n}}\P^\a(V_H)_t
~[(\bar{u}+U)\cdot\nabla)^2,\tilde{\chi}]\P^\a V dx+\{\cdots\}\vspace{2.5mm}\\
=&\dis{-\frac{d}{dt}\int_{{\R}^n}}\frac{1}{2}|(\bar{u}+U)\cdot\nabla\P^\a(V_H)|^2dx
-\dis{\int_{{\R}^n}}(\bar{u}+U)_t\cdot\nabla\P^\a(V_H)(\bar{u}+U)\cdot\nabla\P^\a(V_H)dx\\
&-\dis{\int_{{\R}^n}}\nabla(\bar{u}+U)\cdot\nabla\P^\a(V_H)_t(\bar{u}+U)\cdot\nabla\P^\a(V_H)dx+\dis{\int_{{\R}^n}}\P^\a(V_H)_t
~[(\bar{u}+U)\cdot\nabla)^2,\tilde{\chi}]\P^\a V dx+\{\cdots\}\vspace{2.5mm}\\
=:
&-\dis{\frac{d}{dt}\int_{{\R}^n}}\frac{1}{2}|(\bar{u}+U)\cdot\nabla\P^\a(V_H)|^2dx+N_{1,1}+N_{1,2}+N_{1,3}+\{\cdots\}.
\end{array}
$$
Here and in the subsequent of this section, we use $\{\cdots\}$ to
denote the terms with derivatives of order at most $|\a|+1$. It is
easy to see that
$$
 |N_{1,1}+N_{1,2}+N_{1,3}+\{\cdots\} | \leq C(E_0+{\mathcal
M}^3)(1+t)^{-\frac{n}{2}-(|\a|+3)}.
$$
Then for $N_2$, we have
$$
\begin{array}{rl}
N_2=&\dis{\int_{{\R}^n}}\P^\a(V_H)_t ~\tilde{\chi}\P^\a\triangle
({\mathcal
P}(\bar{\p}, V)V^2)dx\vspace{2.5mm}\\
=&\dis{\int_{{\R}^n}}\P^\a(V_H)_t ~\tilde{\chi}{\mathcal
P}^\prime_V~\P^\a \triangle V~V^2 + 2 \P^\a(V_H)_t
~\tilde{\chi}{\mathcal
P}(\bar{\p}, V)~V~\P^\a\triangle V dx+ \{\cdots\}\vspace{2.5mm}\\
=:& N_{2,1} + N_{2,2} + \{\cdots\}.
\end{array}
$$
By noticing that
$$
\begin{array}{rl}
N_{2,1}=&\dis{\int_{{\R}^n}}\P^\a(V_H)_t ~\tilde{\chi}{\mathcal P}^\prime_V~\P^\a \triangle V~V^2 dx\vspace{2.5mm}\\
=&\dis{\int_{{\R}^n}}\P^\a(V_H)_t ~(V^2{\mathcal P}^\prime_V)~\P^\a
\triangle V_H dx + \dis{\int_{{\R}^n}}\P^\a(V_H)_t
~[\tilde{\chi},{\mathcal P}^\prime_V V^2]~\P^\a
\triangle V_L dx,\vspace{2.5mm}\\
=& -\frac{d}{dt}\dis{\int_{{\R}^n}}({\mathcal P}^\prime_V
V^2)~|\nabla\P^\a V_H|^2 dx+ O_1,
\end{array}
$$
with
$$
O_1\leq \eta \int_{\R^n}\P^\alpha (V_H)|^2dx+C_\eta (E_0+{\mathcal
M}^3)(1+t)^{-\frac{n}{2}-(|\a|+3)}.
$$
Similarly, we have
$$
N_{2,2}=-2\frac{d}{dt}\dis{\int_{{\R}^n}}({\mathcal P}V)~|\nabla\P^\a
V_H|^2 dx+ O_2,
$$
with
$$
O_2\leq \eta \int_{\R^n}|\P^\alpha (V_H)|^2dx+C_\eta (E_0+{\mathcal
M}^3)(1+t)^{-\frac{n}{2}-(|\a|+3)}.
$$

We still need to consider the terms from the expansion of $\P^\a
\mathrm{div}[( \bar{\p}^{\pm}+V^{\pm} )\nabla\varphi], \  0<|\a|\leq
k-1$:
\begin{itemize}
\item If all the derivatives $\P^\a \mathrm{div}$ are taken on $ \bar{\p}^{\pm}+V^{\pm}$, it can be bounded by the a priori assumption and the fact that $\nabla\varphi \in L^{\infty}$.
\item If  all the derivatives $\P^\a \mathrm{div}$ are taken on  $\nabla\varphi  $  then
$$
\P^\a \mathrm{div} \nabla\varphi = \P^\a K,
$$
and recall the good decay properties of $\P^\a K$ for $|\a|\leq k-1$
in Section 3, then we have better decay on these terms.
\item Other terms can be estimated similarly.
\end{itemize}

Again to close the energy estimate, we now use the fact that
$$
\dis{\int_{{\R}^n}} |\nabla_x \P^\a V_H |^2dx\geq
\epsilon\dis{\int_{{\R}^n}} |\P^\a V_H |^2dx.
$$
By integrating (\ref{p3}) and (\ref{p4}) over $[0, t]$ and
multiplying  (\ref{p3}) by some suitably chosen constant
$0<\lambda<1$, the combination of above estimates give (\ref{p13}).
Therefore, we have the following estimates on the high frequency
component.

\begin{thm}
Under the assumption of Theorem 1.1, we have, for $|\a|\leq k-1$,
$$
 \|\P^\a V^{\pm}_H \|_{H^1} +  \|\P^\a (V^{\pm}_H)_t \|_{L^2} \leq
C(E_0+{\mathcal M}^{3/2})
(1+t)^{-\frac{n}{4}-\frac{|\a|+3}{2}},
$$
where $E_0$ and ${\mathcal M}$ are defined in Theorem 4.2 and
(\ref{l19}), respectively.
\end{thm}

\section{Proof of Theorem 1.1.}
In the previous two sections, we  obtain the following
estimates on the low frequency component by using the approximate
Green function and the high frequency component by using the energy
method respectively,
\begin{equation}\label{p6.1}
  \|\P^\a V_L^{\pm} (t) \|_{L^p}\leq C(E_0+{\mathcal M}^2)
(1+t)^{-\frac{n}{2}(1-\frac{1}{p})-\frac{|\a|+1}{2}},~~|\a|\leq k,
\end{equation}
and
\begin{equation}\label{p6.2}
  \|\P^\a V_H^{\pm} \|_{H^1} +  \|\P^\a (V_H^{\pm})_t \|_{L^2}\leq
C(E_0+{\mathcal
M}^{3/2})(1+t)^{-\frac{n}{4}-\frac{|\a|+3}{2}},~~|\a|\leq k-1.
\end{equation}
It remains to combine (\ref{p6.1}) and (\ref{p6.2}) to close the a priori
assumption (\ref{l19}).

Firstly, by taking $p=2$ in (\ref{p6.1}) and combining with (\ref{p6.2}), we have
$$
 \|\P^\a V^{\pm}(t) \|_{L^2}\leq C(E_0+{\mathcal
M}^{3/2})(1+t)^{-\frac{n}{4}-\frac{|\a|+1}{2}},~~|\a|\leq k.
$$
Next, by using the Sobolev embedding theorem, from (\ref{p6.2}), we have, for
$|\a|\leq k-2$,
$$
\begin{array}{rl}
 \|\P^\a V_H^{\pm}(t) \|_{L^\infty} \leq & \|\P^\a V_H^{\pm}(t) \|_{H^2}
\leq  \|\P^\a V_H^{\pm}(t) \|_{L^2} +  \|\nabla\P^\a V_H^{\pm}(t) \|_{H^1}\vspace{2.5mm}\\
\leq &C(E_0+{\mathcal M}^{3/2})(1+t)^{-\frac{n}{4}-\frac{|\a|+3}{2}}.
\end{array}
$$
Moreover, for $n=3$, it holds that
$-\frac{n}{4}-\frac{|\a|+3}{2}\leq-\frac{n}{2}-\frac{|\a|+1}{2}$.
Thus,
\begin{equation}\label{p6.5}
\|\P^\a V_H^{\pm}(t)\|_{L^\infty} \leq C(E_0+{\mathcal
M}^{3/2})(1+t)^{-\frac{n}{2}-\frac{|\a|+1}{2}},~~|\a|\leq k-2.
\end{equation}
Then, the interpolation of (\ref{p6.2}) and (\ref{p6.5}) leads to
\begin{equation}\label{p6.6}
\|\P^\a V_H^{\pm}(t) \|_{L^p}\leq C(E_0+{\mathcal
M}^{3/2})(1+t)^{-\frac{n}{2}(1-\frac{1}{p})-\frac{|\a|+1}{2}},~~|\a|\leq
k-2.
\end{equation}
Combining (\ref{p6.1}) with (\ref{p6.6}) then gives
$$
 \|\P^\a V^{\pm}(t) \|_{L^p}\leq C(E_0+{\mathcal
M}^{3/2})(1+t)^{-\frac{n}{2}(1-\frac{1}{p})-\frac{|\a|+1}{2}},~~|\a|\leq
k-2.
$$

Now, we turn to estimate $U^{\pm}$ by using the equation
(\ref{l16}). Note that
$$
U^{\pm}(x,t)=e^{- t}U^{\pm}(x,0)+\int^t_0e^{-(t-s)}
((\bar{\p}^{\pm})^{-1}\nabla (a^{\pm} V^{\pm})+\bar{H}^{\pm} \pm
\nabla\varphi )(x,s)ds,
$$
and it is easy to check that, for $|\r|\leq
k-2$,
$$
\left\{
\begin{array}{lcr}  \|\P^\r \bar{H}^{\pm}(s) \|_{L^\infty}\leq
 C(E_0+{\mathcal M}^2)
(1+s)^{-(n+1+\frac{|\r|+1}{2}) },\vspace{2.5mm}\\
 \|\P^\r ((\bar{\p}^{\pm})^{-1}\nabla_x(a^{\pm}V^{\pm})(s) \|_{L^\infty}\leq
C(E_\rho+{\mathcal M}_V)
(1+s)^{-\frac{n}{2} -\frac{|\r|+2}{2}}, \vspace{2.5mm}\\
 \|  \P^\r \nabla\varphi  \| _{L^\infty}\leq
 \|  \P^\r K  \| _{H^{1}} \leq C(E_{\rho}+ \mathcal M^{2})
(1+t)^{-{5\over 4} n -2-{|\gamma| \over 2}},
\end{array}
\right.
$$
thus,
\begin{equation}\label{A4}
  \|\P^\r U^{\pm}(t) \|_{L^\infty} \leq C(E_\rho+{\mathcal M}_V+{\mathcal
M}^2)(1+t)^{-\frac{n}{2}-\frac{|\r|+2}{2}},~~|\r|\leq
k-2.
\end{equation}

Next, for the $L^2$-norm, we use energy estimate. Multiply (\ref{l16}) by $\bar \rho^{\pm} U^{\pm}$ and integrate, we have % (omit the sign $\pm$)
\begin{equation}\label{A1}
 \int_{{\R}^n} \bar \rho^{\pm} U^{\pm} U^{\pm}_{t}dx
+ \int_{{\R}^n} \nabla(a^{\pm}V^{\pm}) \cdot U^{\pm}dx
 + \int_{{\R}^n}\bar \rho^{\pm} (U^{\pm})^{2}dx
= \int_{{\R}^n} \bar H^{\pm} \bar \rho^{\pm} U^{\pm}dx \pm
\int_{{\R}^n} \bar \rho^{\pm} \nabla \phi \cdot U^{\pm}dx.
\end{equation}
Note
$$
\int_{{\R}^n} \bar \rho^{\pm} U^{\pm} U^{\pm}_{t} dx= {d\over dt}
\int_{{\R}^n} \bar \rho^{\pm} (U^{\pm})^{2}dx
 - \int_{{\R}^n} \P_{t} \bar \rho^{\pm}  (U^{\pm})^{2}dx,
$$
in which the second term is small, and
\begin{equation}\label{A2}
 | \int_{{\R}^n} \nabla(a^{\pm}V^{\pm}) \cdot U^{\pm}dx|
\leq \varepsilon \int_{{\R}^n}  (U^{\pm})^{2} + C(\varepsilon)
\mathcal M^{2} (1+t)^{-{n\over 2} -(1+1)} ,
\end{equation}
in which the second term has the decay rate of $\|\nabla V^{\pm}\|
^{2}_{L^{2}}$, (which is the key observation that $U^{\pm}$ has
better decay than $V^{\pm}$).

The first term on the right hand side of (\ref{A1}), the  nonlinear
term, can be estimated as in $K$, which has better decay than
(\ref{A2}). For the last term with $\nabla \varphi$, we have
\begin{eqnarray*}
|\int_{{\R}^n} \bar \rho^{\pm} \nabla \phi \cdot U^{\pm}dx| \leq \|
\bar \rho^{\pm} \|_{L^{3}} \|  \nabla \phi   \|_{L^{6}} \|   U^{\pm}
\|_{L^{2}} &\leq&  \| \bar \rho^{\pm} \|_{L^{2}}^{1/2}  \| \bar
\rho^{\pm} \|_{L^{6}}^{1/2}
\| E\|_{L^{2}} \|   U^{\pm}   \|_{L^{2}}\\
&\leq& \varepsilon  \|   U^{\pm}   \|_{L^{2}}^{2} + C(\varepsilon)
\mathcal M^{2} (1+t)^{-{5\over 4}n -2-{1\over 4}}.
\end{eqnarray*}

Perform same estimates for $\P^{\alpha}(|\alpha|\leq k)$, and use
the similar argument as for $V^{\pm}$, we can get
\begin{equation}\label{A3}
 \| \P^{\alpha} U^{\pm} \|_{L^{2}} \leq \mathcal M
 (1+t)^{-{n\over 4 } - {|\alpha|+2 \over 2}}, \hspace{3mm} |\alpha|\leq k.
\end{equation}

Interpolate (\ref{A4}) and (\ref{A3}) to have
$$%\begin{equation}
 \| \P^{\alpha} U^{\pm} \|_{L^{p}} \leq \mathcal M
 (1+t)^{-{n\over 2 }(1-{1\over p}) - {|\alpha|+2 \over 2}}, \hspace{3mm} |\alpha|\leq k-2.
$$%\end{equation}
Then combine the above estimates to get

\begin{thm}
Under the assumption of Theorem 1.1, if the initial data
$(V^{\pm}_0, U^{\pm}_0)$ satisfies that
$$
\begin{array}{rl}
 |\p_+-\p_- |+ \|\nu^{\pm}(\cdot, 0) \|_{L^2\cap
L^1}+ \|\nu^{\pm}_t(\cdot, 0) \|_{L^2} + \|V^{\pm}_0 \|_{H^k\cup L^1} + \|V^{\pm}_t(0)
\|_{H^{k-1}} +
 \|U^{\pm}_0 \|_{H^k}\leq\epsilon_0,
\end{array}
$$
where $\epsilon_0>0$ is a small constant, then there exists a unique
global classical solution $(V^{\pm}, U^{\pm})\in
C([0,\infty),H^{k})\cap C^1((0,\infty), H^{k-1})$ to \eqref{l13}.
Moreover, we have
$$
 \left\{ \begin{array}{rl}
 \|\P^\r_xV^{\pm} \|_{L^p} \leq &C(1+t)^{-\frac{n}{2}(1-\frac{1}{p})-\frac{|\r|+1}{2}}, ~~ |\r|\leq k-2, \vspace{2.5mm}\\

 \|\P^\r_xV^{\pm} \|_{L^2} \leq
&C(1+t)^{-\frac{n}{4}-\frac{|\r|+1}{2}}, ~~ |\r|= k-1,~k,\vspace{2.5mm}\\

 \|\P^\r_xU^{\pm} \|_{L^p} \leq
&C(1+t)^{-\frac{n}{2}(1-\frac{1}{p})-\frac{|\r|+2}{2}}, ~~|r|\leq k-2, \vspace{2.5mm}\\

 \|\P^\r_xU^{\pm} \|_{L^2} \leq
&C(1+t)^{-\frac{n}{4}-\frac{k+1}{2}}, ~~ |\r|= k-1,~k.
\end{array} \right.
$$
\end{thm}
This theorem implies  (\ref{l19}) then closed the a priori
assumption, and then it yields the main results (i) and (ii) in
Theorem 1.1.

Before concluding this paper, we point out that even though the
above discussion is for the space dimension  $n=3$, other higher
dimensional cases can be considered similarly.

\

\noindent {\bf Acknowledgements:} The research of Li is partially
supported by the National Science Foundation of China (Grant No.
11171223), the Ph.D. Program Foundation of Ministry of Education of
China (Grant No. 20133127110007) and the Innovation Program of
Shanghai Municipal Education Commission (Grant No. 13ZZ109).
The research of Liao is partially supported by National Natural Science Foundation of China (Nos.11171211, 11301182), Science and Technology commission of Shanghai Municipality (No. 13ZR1453400).

\end{document}